\documentclass[a4paper,fleqn]{cas-sc}
\usepackage{graphicx}
\usepackage[numbers]{natbib}
\usepackage{epstopdf}
\usepackage{float}
\usepackage{caption}
\usepackage[ruled,vlined]{algorithm2e}
\usepackage{xspace}
\usepackage{xcolor}
\usepackage{tikz}
\usetikzlibrary{calc}
\usetikzlibrary{shapes.arrows}
\usetikzlibrary{positioning,fit,backgrounds}
\usetikzlibrary{positioning,calc}
\usetikzlibrary{decorations.pathreplacing}
\usetikzlibrary{shapes.geometric,shapes.arrows,decorations.pathmorphing}
\usetikzlibrary{snakes}
\usetikzlibrary{matrix,chains,scopes,positioning,arrows,fit}

\def\tsc#1{\csdef{#1}{\textsc{\lowercase{#1}}\xspace}}
\tsc{WGM}
\tsc{QE}
\tsc{EP}
\tsc{PMS}
\tsc{BEC}
\tsc{DE}
\usepackage{float}
\usepackage[ruled,vlined]{algorithm2e}
\usepackage{xspace}
\usepackage{xcolor}
\usepackage{tikz}
\usetikzlibrary{calc}
\usetikzlibrary{shapes.arrows}
\usetikzlibrary{positioning,fit,backgrounds}
\usetikzlibrary{positioning,calc}
\usetikzlibrary{decorations.pathreplacing}
\usetikzlibrary{shapes.geometric,shapes.arrows,decorations.pathmorphing}
\usetikzlibrary{matrix,chains,scopes,positioning,arrows,fit}

\begin{document}
\let\WriteBookmarks\relax
\def\floatpagepagefraction{1}
\def\textpagefraction{.001}

\shorttitle{Petascale Brownian dynamics simulations of highly resolved polymer chains with hydrodynamic interactions using modern GPUs}
\shortauthors{Venkata Siva Krishna et~al.}

\title [mode = title]{Petascale Brownian dynamics simulations of highly resolved polymer chains with hydrodynamic interactions using modern GPUs}                      

\author{Venkata Siva Krishna}
\fnmark[1]                
                        
\author{Praphul Kumar}
\fnmark[1]                          

\author{Bharatkumar Sharma}
\fnmark[2]                          

\author{Indranil Saha Dalal}
\cormark[1]
\fnmark[1]

\address{$^1$Department of Chemical Engineering, Indian Institute of Technology Kanpur, Kanpur-208016, India \\ $^2$NVIDIA Graphics Pvt Ltd, Bangalore, Bangalore-560045, India}

\cortext[cor1]{Corresponding author: I. Saha Dalal, Email: indrasd@iitk.ac.in \\ V.S.K and P.K. contributed equally to the study }

\begin{abstract}
Brownian dynamics simulations of fairly long, highly detailed polymer chains, at the resolution of a single Kuhn step, remains computationally 
prohibitive even on the modern processors. This is especially true when the beads on the chain experience hydrodynamic interactions (HI), which 
requires the usage of methods like Cholesky decomposition of large matrices at every timestep. In this study, we perform Petascale BD simulations, with HI, of fairly long and highly resolved polymer chains on modern GPUs. Our results clearly highlight the inadequacies of the typical models that use beads connected by springs. In this manuscript, firstly, we present the details of a highly scalable, parallel hybrid code implemented on a GPU for 
BD simulations of chains resolved to a single Kuhn step. In this hybrid code using CUDA and MPI, we have incorporated HI using the Cholesky 
decomposition method. Next, we validate the GPU implementation extensively with theoretical expectations for polymer chains at equilibrium and in 
flow with results in the absence of HI. Further, our results in flow with HI show significantly different temporal variations of stretch, in both startup 
extensional and shear flows, relative to the conventional bead-spring models. In all cases investigated, the ensemble averaged chain stretch is much 
lower than bead-spring predictions. Also, quite remarkably, our GPU implementation shows a scaling of $\sim$$N^{1.2}$ and $\sim$$N^{2.2}$ of the computational times for shorter and longer chains in the most modern available GPU, respectively, which is significantly lower than the theoretically expected $\sim$$N^{3}$. We expect our methods and results to pave the way for further analysis of polymer physics in flow fields, with long and highly detailed chain models. 
\end{abstract}

\ExplSyntaxOn
\keys_set:nn { stm / mktitle } { nologo }
\ExplSyntaxOff
\maketitle

\section{Introduction}
The study of dilute polymer solutions is of utmost importance for various natural and industrially relevant processes. Thus, it has been investigated thoroughly over the past few decades by theory, experiments and simulations. For all these studies, the foundation is set by the seminal work of Rouse \cite{rouse1953theory}, which decomposed
the dynamics at equilibrium over normal modes. The calculations were further refined by Zimm \cite{zimm1956dynamics}, by adding hydrodynamic interactions (HI), which brought the results closer to experiments. Note, due to the diluteness criterion, it can be assumed that the different polymer chains are sufficiently far away and does not interact with one another. Hence, the analysis of Rouse and Zimm, where a single chain is considered, is pretty successful. This is the basis of single-chain Brownian dynamics (BD) simulations, which is currently the primary computational method currently for the study of dilute polymer solutions. In this, a polymer chain is modelled as a series of beads connected by springs, where the spring law varies according to the situation, as discussed later. In BD simulations, it is customary to replace the solvent molecules by a continuum. Note, the BD simulation predictions show good agreement with experimental studies - both rheological \cite{lee1997molecular,lee1999flow,bossart1997orientation} and microscopy of single chains \cite{smith1999single,hur2000brownian,schroeder2005characteristic}. This proves the validity of BD simulations, which can then be used further to study the dynamics of polymer chains in complicated scenarios and flow fields, which may occur in a variety of situations in nature or in industry.

As discussed earlier, in BD simulations, a single polymer chain consists of a series of beads and springs, with the solvent molecules being replaced by a continuum. These springs (connectors between the beads) can take various forms, depending on the chain discretization level that we wish to attain. For even a very short chain of a few Kuhn steps, retaining all atomistic-level details would entail a simulation of hundreds to thousands of atoms. Thus, a first-level simplification is to model the chain at the level of one single Kuhn step, which represents a rigid, flexibly rotating unit in the chain. This is perhaps the most resolved chain that one can expect in BD simulations, since retaining further details would also require the solvent-level details to be present, for consistency. However, the computational cost is prohibitive, since for any practically relevant problem, there are several tens of thousands of beads on the chain. Note, several such bead-rod simulations have been performed with extremely few rods \cite{doyle1997dynamic,petera1999brownian,shaqfeh2005dynamics}. Recently, BD simulations have been performed with 500 rods without HI \cite{dalal2012multiple}, but with much lower number of rods when HI is included \cite{dalal2014effects}.

The necessity of further simplifications led to the bead-spring models, which have been hugely popular. These springs are supposed to mimic the dynamics of a large number of Kuhn steps, thereby reducing the number of beads required to model polymer chains. In modern day applications, the spring laws of the Cohen-Padé approximation \cite{cohen1991pade} and Underhill-Doyle have been used \cite{underhill2004coarse}. Although these coarse-grained bead-spring models have shown reasonable success, recent simulations do highlight some inconsistencies in the results for steady shear flow \cite{dalal2012multiple}. Even the successive fine graining scheme \cite{PHAM20089}, using bead-spring simulations to predict the bead-rod chain properties, show some differences at high shear rates. A complete comparison of such models for various flow fields, both at steady state and startup, is yet to be performed.

Thus, for the maximum possible accuracy in BD simulations, there is a need to simulate sufficiently long chains using the bead-rod model, with HI being active. As noted, even with the increase in computing power over the past decades, simulations of a maximum of 100-200 bead-rod chain is only possible in the modern day CPU, when HI is active. This owes to the nature of the algorithm with HI, which involves the Cholesky decomposition of a large matrix at every time step of iteration. With this, the computational requirements grow as $N^3$, where $N$ is the number of beads on the chain. Thus, for even a modest $N$, the compute cost becomes impractical for even the most powerful CPU available currently. To overcome this, techniques like the Chebyshev polynomial approximation \cite{fixman1986construction}, Krylov subspaces \cite{ando2012krylov} and truncated expansion approximation (TEA) \cite{geyer2009n} have been proposed to replace the Cholesky decomposition. To compare their efficiencies, Amir et al. \cite{saadat2014computationally} performed simulations of a bead-spring chain with the semi-implicit predictor-corrector method, from which the Krylov subspace emerged to be the best possible algorithm. However, even with all these sophistications, although the scaling exponent is reduced from 3 (as in Cholesky decomposition), the computational cost still remains prohibitive for a few hundred beads.

However, there exist applications where massive speedups, of two orders of magnitudes or more, have been obtained by harnessing the power of GPU, using CUDA programming. These include methods like MD simulations, where the equations of motion are solved for many particles at every iteration. Taking inspiration from this, we aim to develop and evaluate the performance of an algorithm of BD simulations on the modern-day GPU. As we discuss in this manuscript, we have obtained a highly scalable, parallel, GPU implementation of BD simulations of a bead-rod chain, with HI being active. To best of our knowledge, for the first time, we are able to perform Petascale BD simulations for long bead-rod chains, especially with imposed flow fields, with the number of beads as high as 500 and HI being active, by performing Cholesky decomposition at every time step. Significantly, for this implementation, we are also able to obtain lower scaling exponents (close to 2) on the computational time, even while using the conventional algorithm, while using modern GPUs. Additionally, our study highlights the inconsistencies in the predictions obtained from the coarser bead-spring models in the flow fields, especially when HI is active. 

Here, we note that there exist two open-source software packages, developed about a decade back, which claim to perform BD simulations of macromolecules in homogeneous flows and external electric fields. These are Simuflex \cite{de2009simuflex} and BD-BOX \cite{dlugosz2011brownian}. BD-BOX is especially designed for simulations of either multicomponent systems containing a number of different molecular species or a single molecule. It runs on different architectures using multiple programming models i.e., OpenMP for shared memory systems, MPI for distributed memory systems and CUDA for GPUs. It implements both TEA and the Cholesky decomposition techniques for incorporating HI. Likewise, SIMUFLEX consists of two programs\textcolor{red}{:} BROWFLEX and ANAFLEX, where Browflex generates the trajectory of bead-spring chains using BD simulations. Anaflex computes the properties based on the trajectories generated by Browflex. This code runs on multi-core servers or clusters and implements both the Cholesky decomposition and the Fixman technique \cite{fixman1986construction} for performing multiscale DNA simulations. In these simulations, the computations related to the HI tensor is performed only after several steps, to reduce the computational cost. However, to the best of our knowledge, neither did these packages highlight the accuracy of their implementation by comparing with known results (in flow and without flow), nor did any major study involving the dynamics of polymer chains in dilute solutions use them extensively. Furthermore, BD-BOX was developed based on GPU architectures that are obsolete now. Thus, there is a definite need for optimal implementation of BD simulations on modern GPUs, especially with highly resolved chain models.

Apart from the studies mentioned, a few more investigations are noted here for simulations with HI. Teijeiro et al. \cite{teijeiro2013parallel} provided an elaborate description of the algorithm for coding of the Cholesky and Fixman method for BD simulations of particles, using OpenMP and UPC (Unified Parallel C). Jendrejack et al. \cite{jendrejack2000hydrodynamic} simulated a bead-spring chain of upto 125 units, using a weak first order semi-implicit scheme by applying both Fixman and Cholesky technique, where EV is also considered. Mahdy et al. \cite{moghani2017computationally} have simulated a bead-rod chain with HI and EV by developing several algorithms – the combined Picard-conjugate gradient method, implementation of the Barnes and Hut multipole method within the same and the stress calculations. They also studied the effects of  strong uniaxial extensional flows on a bead-rod chain using the Krylov method.

Thus, we note a lack of efforts towards a parallel implementation of bead-rod simulations, using CUDA and the latest GPU architectures. The existing approach (BD-BOX) was built upon now-obsolete architectures, experienced no usage and is not actively maintained. In fact, simulations of bead-rod chains with HI, where the number of beads run in hundreds, are rare. This sets the importance of this study, where we use the conventional Cholesky decomposition technique to simulate a bead-rod chain with HI, using a hybrid code, which uses CUDA and MPI. Unlike previous approaches, we verify the accuracy of the GPU code by comparing the scaling results from equilibrium simulations with the theoretical results. In the process, we also showcase the following significant features of the new development:
\begin{enumerate}
    \item Impact of using the latest GPU architecture like Volta and Ampere, in a dense system like DGX, which has 8 GPU per node
    \item Scaling efficiency across multinode multi-GPU environment
\end{enumerate}
Additionally, our results also highlight the importance of using such highly resolved chain models (to a single Kuhn step) for applications involving flow fields. The chain dynamics show significant differences, especially in the presence of HI, between the bead-rod chain simulations (which are possible on GPU) and the coarser bead-spring simulations, even when the parameters are calculated using existing procedures. In this study, we highlight both these aspects of our GPU implementation - the computational speed and efficiency as well as the impact in terms of the physical predictions. 

This manuscript is organized in the following manner. We give a brief introduction to GP-GPU computing and CUDA in section 2. In section 3, we briefly discuss the theoretical background and numerical scheme required for the simulations. The parallel implementation of the code is discussed using the NVIDIA Nsight tools in section 4. The simulation results for a bead-rod chain at equilibrium and comparison of bead-rod and bead-spring models in the presence of flow with and without HI, are discussed in section 5. Finally, the key findings are summarized in section 6.

\section{GP-GPU Computing and CUDA}
GPUs are designed as high throughput architectures, allowing developers to launch thousands of threads, thereby accelerating simulations. Each and every thread executes the same instruction across multiple data elements, commonly referred to as Single Instruction Multiple Data (SIMD) parallel execution. GPU comprises of multi-processors known as streaming multiprocessors (SM), which contain multiple cores. With the release of every newer generation of GPU architecture, the number of cores and the memory bandwidth increases. For example, the older generation V100 GPU has 84 SMs, each having 32 cores for FP32 computations, and also has 32 GB of on-chip memory with a memory bandwidth of 900 GB/s. The latest GPU architecture A100 has 108 SMs, each with 64 cores for FP32 computations, and the 80 GB of memory bandwidth increased to 1.2 TB/sec. A GPU ecosystem provides multiple approaches for writing parallel programs. Directive-based approaches like OpenACC \cite{openacc} enables vendor-independent and  higher-level code. In this, a domain scientist spends less effort in understanding the GPU architecture, but may end up gaining only a fraction in terms of the overall performance. For our application, we have selected CUDA C as an approach to get the maximum performance on the latest GPU architectures like V100. Wherever possible, we have leveraged the CUDA libraries, which are optimized by NVIDIA for different GPU architectures.

The CUDA ecosystem provides various tools, and the ones that we have used for our application are Nsight System and Nsight Compute. Nsight System and Nsight Compute are part of the Nsight product family from NVIDIA. Nsight System helps us to analyze the application algorithm on a system-wide basis, while Nsight Compute enables us to optimize the individual CUDA kernels, thereby extracting maximum performance from GPU. 
\section{Theoretical background}
A dilute polymer solution can be modelled by a single polymer chain immersed in solvent molecules, since the inter-chain interactions can be neglected. Currently, it is typical to study the dynamics of polymer chains in dilute solutions by Brownian dynamics simulations of a single chain, following several earlier investigations \cite{dalal2012multiple,dalal2014effects,hsieh2003modeling,larson2005rheology}. In this method, the chain is modelled by a series of beads numbered from 0 to $N$, connected by springs which are numbered from 1 to $N$, while the solvent molecules are replaced by a continuum. Each bead on the chain experiences forces due to the connectors (with nearby beads), drag force due to the relative motion of the bead and solvent and a random Brownian force exerted by the solvent on each bead, owing to thermal fluctuations. Also, it is usual to neglect the inertia in Brownian dynamics (BD). Thus, the equation of motion of each bead in absence of excluded volume and hydrodynamic interactions, is given as:  
\begin{equation}
\frac{d\vec{r}_{i}^{\ast}}{dt^{\ast}}=\vec{F}_{i}^{Br\ast}+\vec {F}_{i}^{Flow \ast}+\bigg(\vec{F}_{i+1}^{S\ast}-\vec{F}_{i}^{S\ast}\bigg)
\label{FDeq}
\end{equation}
 Here, $\vec{r}_{i}^{\ast}$ is the position vector of the $i^{th}$ bead, $\vec{F}_{i}^{Br\ast}$ is the Brownian force acting on $i^{th}$ bead, $\vec{F}_{i}^{Flow\ast}$ is the force on a bead due to the imposed flow field and $\vec{F}_{i}^{S\ast}$ and $\vec{F}_{i+1}^{S\ast}$ denote the forces due to the $i^{th}$ and $(i+1)^{th}$ springs, respectively. The superscript ‘$\ast$’ in Eqn. \ref{FDeq} denotes that the variables have been non-dimensionalized by a well-known procedure described in detail in the previous studies \cite{dalal2012multiple,hsieh2003modeling}. Here, the Brownian force is calculated as \cite{larson2005rheology}:
\vspace{1 mm}
\begin{equation}
\vec{F}_{i}^{Br\ast}=\sqrt\frac{6}{\Delta{t}^{\ast}} \vec{n}_{i}    
\end{equation}
\vspace{1 mm}
where $\vec{{n}}_{{i}}$ is a vector, each of whose components is a random number that is uniformly distributed between -1 and 1. Here, $\Delta{t}^{\ast}$ is the simulation timestep size.  In our simulations, we have considered the bead-rod model. For this model, we have used stiff Fraenkel springs, which mimics a rigid rod, following recent studies \cite{dalal2012multiple,kumar2022fraenkel}. The spring force is given as:
\begin{equation}
\vec{F}_{i}^{S\ast}=K^{\ast}\frac{(\mid\vec{Q}_{i}^{\ast}\mid-1)}{\mid \vec{Q}_{i}^{\ast}\mid}\vec{Q}_{i}^{\ast}
\end{equation}
where $K^{\ast}$ represents the stiffness of the spring and $\vec{Q}_{i}^{\ast}$ is the link vector.
The results of these bead-rod simulations are compared with coarse-grained bead-spring simulations. For the bead-spring model, we consider the Cohen-Padé approximation \cite{cohen1991pade} to the inverse Langevin function, which is given as:
\begin{equation}
\left|\vec{F}_{i}^{S\ast}\right|=\frac{\alpha \hat r-\beta \hat r^{3}}{1-\hat r^{2}}
\end{equation}
where $\hat r$ is the fractional extension of the spring (i.e spring length scaled by $\nu b_{k}$, where $\nu$ denotes the number of Kuhn steps mimicked by each spring), $\alpha$ and $\beta$ are parameters that depend on the spring law. For this approximation, the values of  $\alpha$ and $\beta$ are 3 and 1, respectively.    

The flow force is given as:
\begin{equation}
\vec{F}_{i}^{Flow\ast}= \overleftrightarrow{\kappa}^{\ast}\cdot\vec{r}_{i}^{\ast}
\end{equation}
where $\overleftrightarrow{\kappa}^{\ast}=[\nabla \vec{{v}}^{\ast}]^{T}$, where $\nabla$ is gradient operator and $\vec{{v}}^{\ast}$ is the velocity field of the flowing solvent.

In the case of uniaxial extensional flow:

$\overleftrightarrow{\kappa}^{\ast}=\begin{bmatrix}
 \dot{\epsilon} & 0 & 0\\
0 & \frac{-\dot{\epsilon}}{2} & 0\\
0 & 0 & \frac{-\dot{\epsilon}}{2}
\end{bmatrix}$

Here, $\dot{\epsilon}$ represents the extensional rate. 
In the case of steady shear flow:

$\overleftrightarrow{\kappa}^{\ast}=\begin{bmatrix}
0 & \dot{\gamma}  & 0\\
0 & 0 & 0\\
0 & 0 & 0
\end{bmatrix}$

Here, $\dot{\gamma}$ represents the shear rate. 

\textit{Bead-rod model with Hydrodynamic Interactions}:
When HI is active, the equation of motion of the $i^{th}$ bead changes as follows \cite{hsieh2003modeling}: 
\begin{equation}
\frac{d\vec{r}_{i}^{\ast}}{dt^{\ast}}=\overleftrightarrow{\kappa}^{\ast}\cdot\vec{r}_{i}^{\ast}+\sum_{j=0}^{N}\overleftrightarrow{D}_{ij}^{\ast}\cdot\vec{F}^{S\ast}_{j}+\bigg(\sqrt{\frac{6}{\Delta{t}^{\ast}}}\bigg)\sum_{j=0}^{i+1}\overleftrightarrow{\sigma}_{ij}^{\ast}\cdot\vec{n}_{j}
\label{HIeq}
\end{equation}        
where $\vec{F}_{j}^{S \ast}$ is the net spring force acting on the $j^{th}$ bead and $\overleftrightarrow{D}^{\ast}$ is the diffusion tensor, usually taken as the Rotne-Prager-Yamakawa (RPY) tensor \cite{rotne1969variational,yamakawa1972translational} whose components are defined as follows:
 \begin{center}
 \begin{equation}
\overleftrightarrow{D}_{ij}^{\ast}=\overleftrightarrow{I},   i=j
\label{Diffusioneq1}
 \end{equation}
 \end{center}
 \begin{equation}
\overleftrightarrow{D}_{ij}^{\ast}=\frac{3h^{*}}{4R_{ij}}\Bigg\lbrace\Bigg[\bigg(1+\frac{2a^{2}}{3R^{2}_{ij}}\bigg)\overleftrightarrow{I} + \bigg(1-\frac{2a^{2}}{R^{2}_{ij}}\bigg)\frac{\vec{R}_{ij}\vec {R}_{ij}}{R^{2}_{ij}}\Bigg],  R_{{ij}}\geq 2a,i\neq j\Bigg\rbrace
\label{Diffusioneq2}
\end{equation}
\begin{equation}
\overleftrightarrow{D}_{ij}^{\ast}=\frac{3h^{*}}{4R_{ij}}\bigg(\frac{R_{ij}}{2a}\bigg)\Bigg\lbrace\Bigg[\bigg(\frac{8}{3}-\frac{3R_{ij}}{4a}\bigg)\overleftrightarrow{I}+\bigg(\frac{R_{ij}}{4a}\bigg)\frac{\vec{R}_{ij}\vec{R}_{ij}}{R^{2}_{ij}}\Bigg], R_{ij}< 2a, i \neq j\Bigg\rbrace
\label{Diffusioneq3}
\end{equation}
where $a$ is bead radius, $h^{*}$ is the hydrodynamic interaction parameter, $\overleftrightarrow{I}$ is the identity tensor and $\vec{R}_{ij}=\vec{r}_{i}^{\ast}-\vec{r}_{j}^{\ast}$.

The tensor $\overleftrightarrow{\sigma}^{\ast}$ is usually obtained by the Cholesky Decomposition \cite{ermak1978brownian} of $\overleftrightarrow{D}^{\ast}$ as follows:
\begin{equation}
\overleftrightarrow{D}_{ij}^{\ast}=\sum_{k=0}^{n}
\bigg(\overleftrightarrow{\sigma}_{ik}^{\ast}\cdot\overleftrightarrow{\sigma}_{jk}^{\ast}\bigg)
\end{equation}

\textit{ Numerical scheme to find the Link vector with hydrodynamic interactions:} 

We use a semi-implicit scheme to solve the equation of motion for each bead in the chain \cite{hsieh2003modeling}. If we subtract the equation of motion of 
$(i+1)^{th}$ bead from that of the ${i^{th}}$ bead in presence of HI, the following is 
obtained for the $i^{th}$ link vector $\vec{Q}_i^\ast$:
\begin{equation}
\frac{d\vec{Q}_{i}^{\ast}}{dt^{\ast}}=\overleftrightarrow{\kappa}^{\ast}\cdot\vec{Q}_{i}^{\ast}+\sum_{j=0}^{N}\bigg(\overleftrightarrow{D}_{i+1,j}^{\ast}-\overleftrightarrow{D}_{i,j}^{\ast}\bigg)\cdot\vec{F}^{S\ast}_{j}+\bigg(\sqrt{\frac{6}{\Delta{t}^{\ast}}}\bigg)\sum_{j=0}^{i+1}\bigg(\overleftrightarrow{\sigma}^{\ast}_{i+1,j}-\overleftrightarrow{\sigma}^{\ast}_{i,j}\bigg)\cdot\vec{n}_{j}
\label{HIQeq}
\end{equation}
Discretization of Eqn. \ref{HIQeq} for a small time step $\Delta t^\ast$ yields the 
following:
\begin{equation}
\left(\vec{Q}_i^\ast\right)^{ t^\ast+\Delta t^\ast} =\left(\vec{Q}_i^\ast\right)^{ t^\ast}+\Delta t^\ast
\left\{ \overleftrightarrow{\kappa }^\ast\cdot\vec{Q}^{\ast}_{i}+\sum_{j=0}^{N}\bigg(\overleftrightarrow{D}^{\ast}_{i+1,j}-\overleftrightarrow{D}^{\ast}_{i,j}\bigg)\cdot\vec{F}^{S*}_{j}+\sqrt{\frac{6}{\Delta t^{*}}}\sum_{j=0}^{i+1}\bigg(\overleftrightarrow{\sigma}^{\ast}_{i+1,j}-\overleftrightarrow{\sigma}^{\ast}_{i,j}\bigg)\cdot\vec{n}_{j}\right\}^{t^\ast}
\label{eqQ}
\end{equation}

In case of the semi-implicit method \cite{hsieh2003modeling}, the spring force of the $i^{th}$ spring (third term on RHS of Eqn. \ref{eqQ}) is evaluated at $t^\ast+\Delta t^\ast$ instead of $t^\ast$. Therefore, this term is brought to LHS and the total term on LHS is equated to a new variable as shown in Eqn. \ref{eqQ1}: 

\begin{equation}
 \vec{Y}_i^{t^\ast+\Delta t^\ast}={\vec{Q}_i^{\ast t^\ast+\Delta t^\ast}} + \Delta t^\ast \left\{ \bigg(\overleftrightarrow{D}^{\ast}_{i+1,i+1}+\overleftrightarrow{D}^{\ast}_{i,i}\bigg)\cdot\vec{F}^{S*}_{i}\right\}^{t^*+\Delta t^*}
 \label{eqQ1}
\end{equation}
This implies that Eqn. \ref{eqQ} can be written as shown below in Eqn. \ref{eqQ2}:
\begin{equation}\label{eqQ2}
\vec{Y}_i^{t^\ast+\Delta t^\ast}=\left(\vec{Q}_i^\ast\right)^{ t^\ast}+\Delta t^\ast
\left\{ \overleftrightarrow{\kappa }^\ast\cdot\vec{Q}^{\ast}_{i}+\sum_{j=0}^{N}\bigg(\overleftrightarrow{D}^{\ast}_{i+1,j}-\overleftrightarrow{D}^{\ast}_{i,j}\bigg)\cdot\vec{F}^{s*}_{j}+2\vec{F}^{s*}_{i}+\sqrt{\frac{6}{\Delta t^{*}}}\sum_{j=0}^{i+1}\bigg(\overleftrightarrow{\sigma}^{\ast}_{i+1,j}-\overleftrightarrow{\sigma}^{\ast}_{i,j}\bigg)\cdot\vec{n}_{j}\right\}^{t^\ast}
\end{equation}

We can further expand Eqn.\ref{eqQ1} by using the Fraenkel spring law to obtain the following: 

\begin{equation}\label{eq:23}
 \vec{Y}_i^{t^\ast+\Delta t^\ast}={\vec{Q}_i^{\ast t^\ast+\Delta t^\ast}} \left[ 
1+2\Delta t^\ast\dfrac{K^\ast\left(\left| {\vec{Q}_i^{\ast t^\ast+\Delta t^\ast}}\right|-1\right)   }{\left| {\vec{Q}_i^{\ast t^\ast+\Delta t^\ast}}\right|}\right] 
\end{equation}

Now, taking the moduli on both sides of Eq. \ref{eq:23}, we arrive at the following form:
\begin{equation}\label{eq:24}
\left| \vec{Y}_i^{t^\ast+\Delta t^\ast} \right|=\left|{\vec{Q}_i^{\ast t^\ast+\Delta t^\ast}}\right| \left[ 
1+2\Delta t^\ast\dfrac{K^\ast\left(\left| {\vec{Q}_i^{\ast t^\ast+\Delta t^\ast}}\right|-1\right)   }{\left| {\vec{Q}_i^{\ast t^\ast+\Delta t^\ast}}\right|}\right] 
\end{equation}

This equation can be further simplified: 
\begin{equation}\label{eq:25}
\left| {\vec{Q}_i^{\ast t^\ast+\Delta t^\ast}}\right|=\dfrac{2\Delta t^\ast K^\ast+\left| 
	\vec{Y}_i^{t^\ast+\Delta t^\ast}\right|}{2\Delta t^\ast K^\ast+1}
\end{equation}
The steps followed to compute the link vectors at ${t^\ast+\Delta t^\ast}$ are given as follows:
\begin{itemize}
    \item Solve Eqn.\ref{eqQ2} to obtain $\vec{Y}_i^{t^\ast+\Delta t^\ast}$ and calculate its moduli. 
    \item  Substitute the moduli of $\vec{Y}_i^{t^\ast+\Delta t^\ast}$ in Eqn. \ref{eq:25} to obtain the moduli of link vectors $\left|{\vec{Q}_i^{t^\ast+\Delta t^\ast}}\right|$.
    \item Once the moduli of link vectors $\left| {\vec{Q}_i^{t^\ast+\Delta t^\ast}}\right|$ and $\vec{Y}_i^{t^\ast+\Delta t^\ast}$ are known, substitute these in Eqn. \ref{eq:23} to calculate all components of the link vectors.
    \end{itemize}

\section{Simulation Methodology}
With the inclusion of HI, the computations require the calculation of the Diffusion tensor and its Cholesky decomposition at every timestep. For a short chain, the serial implementation is feasible on the modern CPUs, since the computational time requirements remain practical. However, with an increase in chain length (or $N$), the cost of the Cholesky decomposition step increases as $N^3$. Hence, the computational times required for the serial implementation for even moderately long chains, say a few hundred beads, becomes prohibitive. Thus, for performing BD simulations of such long chains with HI, we have used the parallel implementation of the code using CUDA on GPU.

Also, under the imposed flow field conditions, it is necessary to perform averaging over multiple initial configurations for the calculation of any property. To compute such an ensemble average, the same CUDA code needs to be run multiple times using different initial configurations. Instead of running the same code multiple times on a single GPU, the work is distributed across multiple GPUs. For this, an MPI-based approach with CUDA-C is used to adapt the code for a hybrid multi-node CPU-GPU environment. The code was run across nodes having multiple GPUs, each with a different initial configuration. Since there is no communication across the MPI processes, we expect an ideal speed-up across the nodes. The hybrid code is depicted by the flow sheet in Fig. \ref{flowchartserialcode}.
\tikzstyle{startstop} = [rectangle, rounded corners, minimum width=1cm, minimum height=1cm, draw=black, fill=red!30]
\tikzstyle{io} = [trapezium, trapezium left angle=70, trapezium right angle=110, minimum width=1cm, minimum height=1cm, draw=black, fill=blue!30]
\tikzstyle{block} = [rectangle, draw, fill=yellow!50,text width=8em, text centered, rounded corners, minimum height=5em]
\tikzstyle{blockexp} = [ellipse, draw, fill=yellow!50, text centered, rounded corners, minimum width=3em,minimum height=3em]
\tikzstyle{block1} = [rectangle, draw, fill=green!50,text width=8em, text centered, rounded corners, minimum height=5em]
\tikzstyle{blockexp1} = [ellipse, draw, fill=green!50, text centered, rounded corners, minimum width=3em,minimum height=3em]
\tikzstyle{decision} = [diamond, minimum width=0.5cm, minimum height=0.5cm,draw=black, fill=blue!30]
\tikzstyle{arrow} = [thick,->,>=stealth]

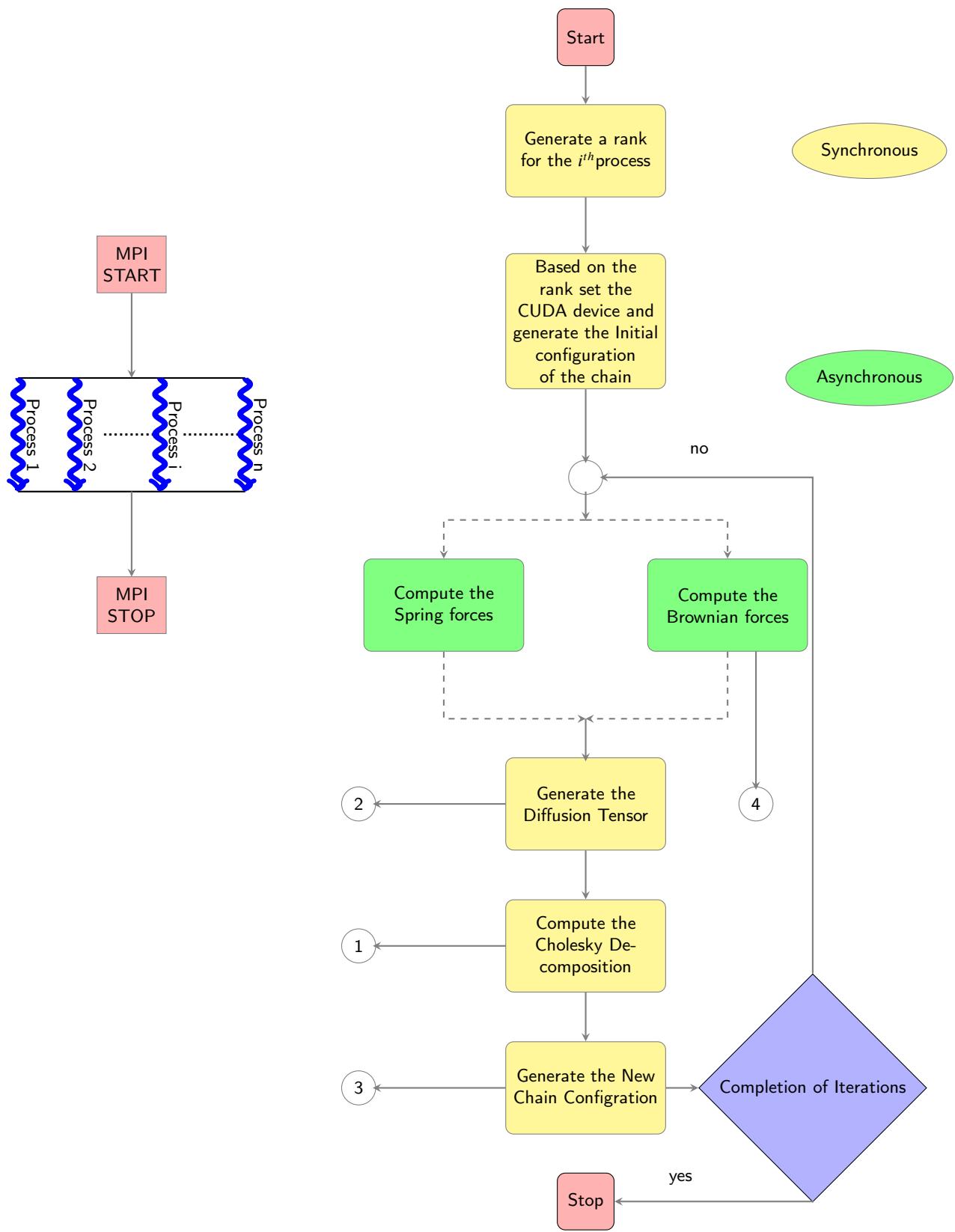
\begin{figure}[hbt!]
\centering
\begin{tikzpicture}
\node[text width=1cm,minimum height=1cm,align=center,fill=red!30] (shape)  at (-8,-4)  [draw] {\color{black}MPI \\START};

\draw[thick,color=black] (-10,-6) -- (-6,-6);
\draw[thick,color=black] (-10,-8) -- (-6,-8);

\draw [thick,->,snake=snake,line width=1mm,color=blue] (-10,-6) -- (-10,-8)
node [midway,sloped,above,color=black]
{Process 1};
\draw [thick,->,snake=snake,line width=1mm,color=blue] (-9,-6) -- (-9,-8)
node [midway,sloped,above,color=black]
{Process 2};
\node at (-6.6,-7.0){$\color{black}\textbf{...........}$};
\node at (-8.0,-7.0){$\color{black}\textbf{...........}$};
\draw [thick,->,snake=snake,line width=1mm,color=blue] (-7.5,-6) -- (-7.5,-8)
node [midway,sloped,above,color=black]
{Process i};
\draw [thick,->,snake=snake,line width=1mm,color=blue] (-6,-6) -- (-6,-8)
node [midway,sloped,above,color=black]
{Process n};
\node[text width=1cm,minimum height=1cm,align=center,fill=red!30] (shape)  at (-8,-10)  [draw] {\color{black}MPI \\STOP};
\draw [arrow] (-8,-4.5) -- (-8,-6);
\draw [arrow] (-8,-8) -- (-8,-9.5);
\node (start) [startstop] {\color{black} Start};
\node (pro1)[block, below of=start,yshift=-1cm] {\color{black}Generate a rank for the $i^{th}$process};
\node (pro2)[block, below of=pro1,yshift=-2cm] {\color{black}Based on the rank set the CUDA device and generate the Initial configuration of the chain};
\draw[arrow] (0,-6.2) -> (0,-7.5);
\draw[arrow] (0,-8.0) -> (0,-8.5);
\draw (0,-7.75) circle (0.3cm);
\draw (-4,-13.5) circle (0.3cm) 
node[color=black] {2};
\draw[arrow] (0,-13.5) -- (-3.75,-13.5);
\draw (-4,-16.0) circle (0.3cm) node[color=black] {1};
\draw[arrow] (0,-16) -- (-3.75,-16);
\draw (-4,-18.5) circle (0.3cm) node[color=black] {3};
\draw[arrow] (0,-18.5) -- (-3.75,-18.5);
\draw (3,-13.5) circle (0.3cm) 
node[color=black] {4};
\draw[arrow] (3,-10) -- (3,-13.25);
\node (pro3) [block1, below of=pro2,xshift=-2.50cm,yshift=-4.0cm] {\color{black}Compute the Spring forces};
\node (pro4) [block1, right of=pro3,xshift=4.0cm] {\color{black}Compute the Brownian forces};
\node (pro5) [block, below of=pro1,yshift=-10.5cm] {\color{black}Generate the Diffusion Tensor};
\draw[dashed,->,>=stealth,thick] (0,-8.5) -| (node cs:name=pro4,anchor=north);
\draw[dashed,->,>=stealth,thick] (0,-8.5) -| (node cs:name=pro3,anchor=north);
\draw[dashed,->,>=stealth,thick] (node cs:name=pro3, anchor=south) |- (0,-12.0);
\draw[dashed,->,>=stealth,thick] (node cs:name=pro4, anchor=south) |- (0,-12.0);
\draw[arrow] (0,-12.0) -> (0,-12.75);
\node (pro6) [block, below of=pro5,yshift=-1.5cm] {\color{black}Compute the Cholesky Decomposition};
\node (pro7) [block, below of=pro6,yshift=-1.5cm] {\color{black}Generate the New Chain Configration};

\node (dec1) [decision, right of=pro7, xshift=3cm] {\color{black}Completion of Iterations};

\node (stop) [startstop, below of=pro7,yshift=-1.0cm] {\color{black}Stop};
\node (pro8) [blockexp, right of=pro1, xshift=4cm] {\color{black}Synchronous};
\node (pro9) [blockexp1, right of=pro4, yshift=4cm, xshift=1.5cm] {\color{black}Asynchronous};
\draw [arrow] (start) -- (pro1);
\draw [arrow] (pro1) -- (pro2);
\draw [arrow] (pro5) -- (pro6);
\draw [arrow] (pro6) -- (pro7);
\draw [arrow] (pro7) -- (dec1);
\draw [arrow] (dec1) |- node[anchor=east, yshift=0.4cm, xshift=-2cm] {\color{black}yes} (stop);
\draw[arrow]  (dec1) |- node[anchor=south, yshift=0.3cm, xshift=-2cm] {\color{black}no} (0.25,-7.75);
\end{tikzpicture}
\caption{A detailed flowchart showing the sequence of the major steps in the parallel implementation of the hybrid BD simulation code with HI.}
\label{flowchartserialcode}    
\end{figure}

The flowchart in Fig. \ref{flowchartserialcode} further explains the flow of the code for BD simulations of dilute polymer solutions running on each single GPU, where we generate the new configuration of the chain by  solving the equations of motion by using the semi-implicit method, which is discussed in details elsewhere \cite{hsieh2003modeling}. Further, the implementation of the CUDA code is analyzed by calculating the computational time requirements for a simulation of 10,000 iterations of a 500 bead-rod chain on a GPU (NVIDIA V100-SXM2-16GB), for which the breakup is shown in Table \ref{final}. These are obtained by profiling the code with the latest Nsight tools (part of NVIDIA HPC SDK), and the steps that consume 99\% of the total time are listed here as well as highlighted in the flow chart.
\begin{table}[width=0.9\linewidth,pos=h]
\centering
\caption{The functions and their execution times in \% that consume the major part of the total time for 10,000 iterations on GPU (Telsa V100-PCIE-32GB) for chain of 500 rods.}
\begin{tabular}{|c|c|c|c|} 
 \hline
 S.No & \textbf{Functions} & \textbf{time}(\%)\\
 \hline
 1) & Cholesky Decomposition  & 89.09\\ 
 2) & Diffusion Tensor       &  2.97\\
 3) & Multiplication and Reduction & 7.8\\
 4) & Generating random numbers &  0.08\\
 \hline
 \end{tabular}
 \label{final}
 \end{table}

 In what follows, we show the parallel pseudocode for these functions on a GPU.

\textbf{Cholesky Decomposition:} To parallelize this step, we use the existing math libraries developed by NVIDIA. Since the code involves the decomposition of a matrix, the API \textit{cuSolverDnDpotrf} is used from the \textit{cuSolver} \cite{cuSolver} library. This is a CUDA library used for decompositions and linear system solutions for both dense and sparse matrices.

\textbf{Scaling of Cholesky Decomposition:-}
 Further, we have studied the scaling of the time consumed by the Cholesky decomposition step as the size of the Diffusion Tensor matrix is increased (this is a function of $N$). This is highly pertinent for this problem, since this step consumes the largest amount of time. For this, we run simulations from which we noted the time taken by the Cholesky decomposition step for 10,000 iterations for different values of ${N}$. Note, for ${N}$ $>=$ 7000, the GPU runs out of memory. Thus, this analysis is restricted upto $N=$ 6399. Currently, we can’t simulate for chains with more than 7000 beads without using the concept of Unified Memory - a feature of the latest GPU and CUDA versions. This will be a part of a future study.
 \begin{figure}[hbt!]
\centering
\includegraphics[scale=0.5]{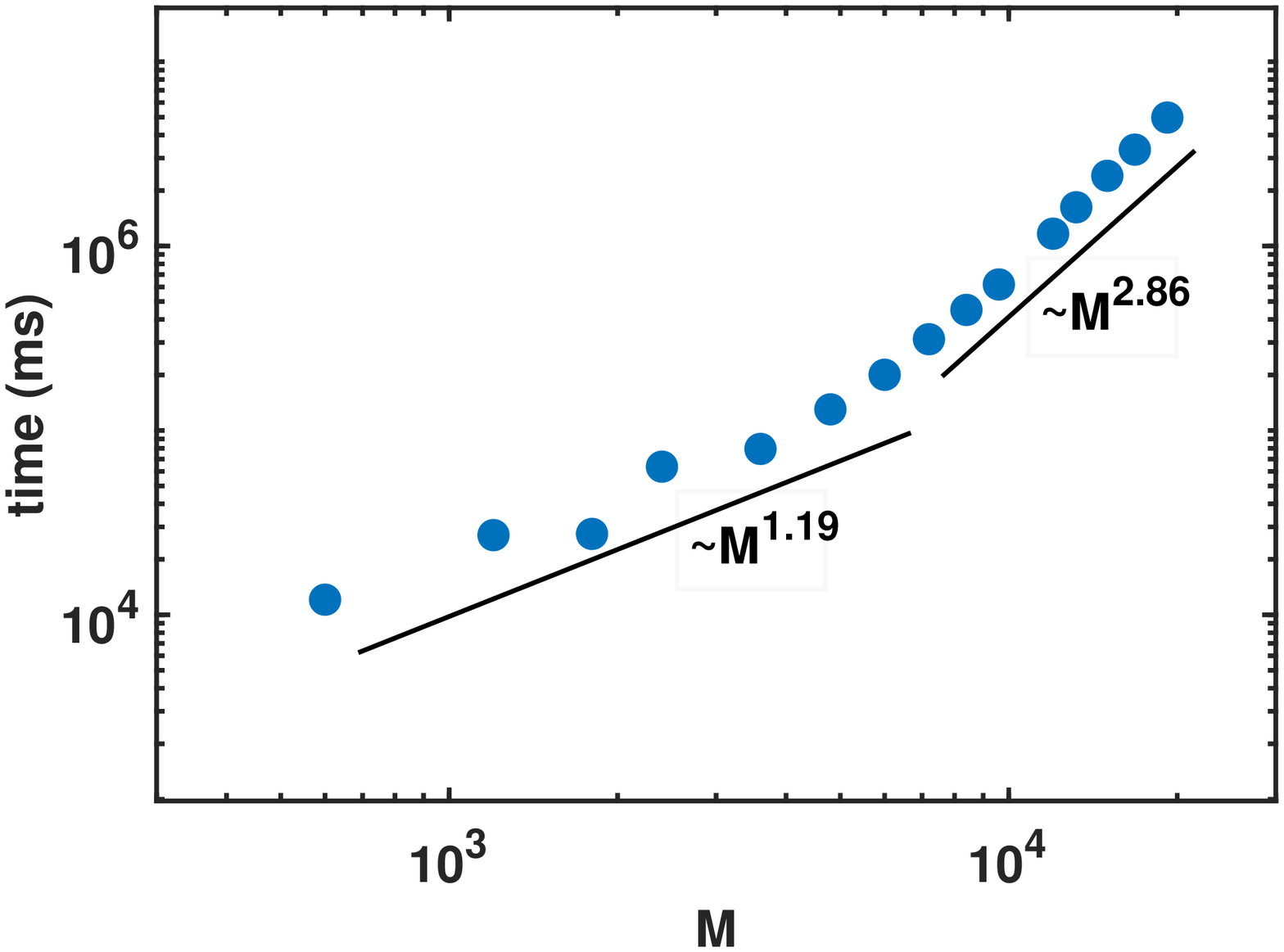}
\caption{Scaling analysis of the time taken by the Cholesky decomposition kernel for 10,000 iterations with the row size of the Diffusion Tensor Matrix (M). The GPU used is NVIDIA V100-SXM2-16GB.}
    \label{scalingcholeskyfig}
\end{figure}

 From Fig. \ref{scalingcholeskyfig}, we observe two different scalings i.e. ${M}^{1.19}$  when the value of $M$ ranges from 600 to 4800 and ${M}^{2.86}$ for larger M. Note, the scaling for lower M is far more desirable than $M^3$, which is theoretically expected, whereas that for larger M is close to the theoretical expectation. Here the variable $M$ represents the row size of the Diffusion Tensor matrix i.e $M$=3*$(N+1)$, which undergoes the Cholesky Decomposition.

To further understand this behaviour, we have performed a load analysis on GPU using Nsight compute by doubling the number of beads, ranging from 200 to 6400 i.e. $N+1$=200,400,800,..,6400. This analysis gives details of the SM utilization  \% and SM memory \% of different kernels that are called by the library function \textit{cuSolverDnDpotrf} during execution. We observe that, as the chain size increases, the kernels that are called by the library function differs. We have considered only the specific kernels that are common for all $N$ and that consume maximum SM utilization  \% and SM memory \% (\textit{volta\_dgemm\_128x64\_upper\_tn kernel}, \textit{trsm\_left\_kernel}) to study the underlying reason behind observing  two different scalings. These are shown in  Tables \ref{kernelutlizationtable} and \ref{kernel2utlizationtable}.

From Table \ref{kernelutlizationtable}, we observe that, by increasing the number of beads from 1600 to 6400,  neither the SM utilization \% nor the SM Memory \% is increased nor it has reached the saturation level. Consequently, the time taken for the execution of the kernel remains largely unchanged. Thus, due to the resources not being utilized completely, the scaling remains much lower than the theoretical expectations (i.e. $M^{3}$) till $N$=1600. The other kernel \textit{volta\_dgemm\_128x64\_upper\_tn} is called by the library function only when we simulate with number of beads greater than 1600. Also, from Table \ref{kernel2utlizationtable} , we do not observe any significant change in  the values of SM utilization \% and the SM Memory \% as we increase the number of beads. However, there is a significant difference in the time taken for the execution of the kernel. From this, we can conclude that for long chain simulations, the GPU computational time is increasing significantly, thereby showing a scaling law close to that for a serial implementation in a CPU. 
\begin{table}[width=0.9\linewidth,pos=h]
\caption{SM Utilization\% , SM Memory\% and time consumed by the \textit{void trsm\_left\_kernel} for evey iteration.}
\centering
\begin{tabular}{|c|c|c|c|} 
 \hline
 \textbf{N+1} &\textbf{SM (Memory)\%}&\textbf{SM Utilization\%} &\textbf{time ($\mu$s)}\\  
 \hline
 200 & 1.76  &	1.16 & 	40.64\\
\hline
400 & 17.39 &	11.36 & 29.76\\
\hline
800 & 44.98 &	29.18 & 31.81 \\
\hline
1600 & 65.72 &	42.46 &	46.98 \\
\hline
3200 & 76.82 &	49.58 &	49.58 \\
\hline
6400 & 73.86 &	47.67 &	47.55 \\
\hline
 \end{tabular}
\label{kernelutlizationtable}
\end{table}
\begin{table}[width=0.9\linewidth,pos=h]
\caption{SM Utilization\% , SM Memory\% and time consumed by \textit{volta\_dgemm\_128x64\_upper\_tn kernel} for every iteration}
\centering
\begin{tabular}{|c|c|c|c|} 
 \hline
 \textbf{N+1} &\textbf{SM (Memory)\%}&\textbf{SM Utilization\%} &\textbf{time (ms)}\\  
 \hline
3200 & 40.76 &	87.6  &	2.07 \\
\hline
6400 & 34.28 &	72.07 &	40.98 \\
\hline
 \end{tabular}
\label{kernel2utlizationtable}
\end{table}

\textbf{Diffusion Tensor and Multiplication} 
To show the working of the Diffusion Tensor and Multiplication kernels, let us define the diffusion tensor for a simple case of a polymer chain with a single rod (\textit{N})  and two beads (\textit{N+1}), as shown in Eqn. \ref{EqnDTn}:

\begin{equation}
\textit{DTn}=\begin{bmatrix}
\overleftrightarrow{D}_{00} & \overleftrightarrow{D}_{01}\\
\overleftrightarrow{{D}}_{10} & \overleftrightarrow{{D}}_{11}
\end{bmatrix}
\label{EqnDTn}
\end{equation}

Each element of this matrix is a tensor in itself. So, we can further expand the first element of this matrix as shown in Eqn. \ref{EqnCompMatrix}:
\begin{equation}
\overleftrightarrow{D}_{00}=\begin{bmatrix}
{D}_{0000} & {D}_{0001} & {D}_{0002} \\ {D}_{0010} & {D}_{0011} & {D}_{0012} \\
{D}_{0020} & {D}_{0021} & {D}_{0022} \\
\end{bmatrix}
\label{EqnCompMatrix}
\end{equation}
\\
After expanding all the elements, we obtain the matrix as shown in Eqn. \ref{EqbFullMatrix}:

\begin{equation}
\textit{DTn}=\begin{bmatrix}
{D}_{0000} & {D}_{0001} & {D}_{0002} & {D}_{0100} & {D}_{0101} & {D}_{0102} \\
{D}_{0010} & {D}_{0011} & {D}_{0012} & {D}_{0110} & {D}_{0111} & {D}_{0112} \\
{D}_{0020} & {{D}}_{0021} & {D}_{0022} &  {D}_{0120} & {D}_{0121} & {D}_{0122} \\ \\
{D}_{1000} & {D}_{1001} & {D}_{1002} & {D}_{1100} & {D}_{1101} & {D}_{1102}\\
{D}_{1010} & {D}_{1011} & {D}_{1012} & {D}_{1110} & {D}_{1111} & {D}_{1112} \\
{D}_{1020} & {D}_{1021} & {D}_{1022} & {D}_{1120} & {D}_{1121} & {D}_{1122} 
\end{bmatrix}
\label{EqbFullMatrix}
\end{equation}

\textit{Diffusion Tensor:} 
 Each component of this tensor \textit{DTn} is computed by Eqns \ref{Diffusioneq1}-\ref{Diffusioneq3} given in the “theoretical background” section. The pseudocode for the parallel implementation is shown in Fig \ref{fig:algo-parallel-DTn}.
 A  total of $(N+1)*(N+1)*ndim*ndim$ number of CUDA threads (this product signifies the total number of elements in the diffusion tensor) are released, making the calculation of each component of the tensor independent of each other.
 Here, $ndim$=3 represents the three spatial dimensions.
 $i$ and $j$ in the pseudocode denote the interaction  between different beads and the variables $k$ and $l$ are used to represent the three spatial coordinates of these beads.
\begin{figure}[hbt!]
    \centering
\begin{algorithm}[H]
    \DontPrintSemicolon
\vspace{1 mm}
    \SetAlgoLined
    \vspace{1mm}
    \SetKwFunction{FMain}{Diffusion Tensor}
    \SetKwProg{Fn}{subroutine}{}{}
        \Fn{\FMain}
     {
           declare tid=blockDimx.x*blockIdx.x+threadIdx.x\\
         \If{$tid\leq(N+1)*(N+1)*ndim*ndim$} 
                      { 
                        Generate i, j, k, l based on tid
                        \\
                         \eIf{i==j} 
                      { 
                        $D_{ij}$ is an identity tensor
                      }
                      {
                        $D_{ij}$ depends on distance between the beads i and j
                      }
                    
                      }
                      
    }
        
    \vspace{1mm}
    \textbf{end subroutine}
    \vspace{1mm}
        
    \caption{Algorithm to calculate the Diffusion tensor}
\end{algorithm}
\caption{Parallel pseudocode for the computation of the  Diffusion tensor.}
    \label{fig:algo-parallel-DTn}
\end{figure}

Now, the difference of the elements of the diffusion tensor are computed i.e. $\overleftrightarrow{D}_{10}-\overleftrightarrow{D}_{00}$ and $\overleftrightarrow{D}_{11}-\overleftrightarrow{D}_{01}$. This new matrix is denoted as the difference tensor, which is shown in Eqn. \ref{EqnDiffMatrix}:
\\
\begin{equation}
\textit{Diff}=\begin{bmatrix}
\overleftrightarrow{{D}}_{{10}} -\overleftrightarrow{{D}}_{{00}} &
\overleftrightarrow{{D}}_{{11}}  -\overleftrightarrow{{D}}_{{01}}
\end{bmatrix}
\label{EqnDiffMatrix}
\end{equation}

The spring force vector is shown in Eqn. \ref{EqnForce}:
\begin{equation}
\vec{F}^{S}=\begin{bmatrix}
 {F}_{1}\\ {{F}}_{2} \\ {F}_{3} \\ -{F}_{1} \\ -{F}_{2} \\ -{F}_{3} 
\end{bmatrix}
\label{EqnForce}
\end{equation}
Here ${F}_{1}, {{F}}_{2}, {F}_{3}$ are the components of the spring force acting on bead 1 and $-{F}_{1}, -{F}_{2}, -{F}_{3}$ are the components of the spring force acting on bead 2, respectively. 

\textit{Multiplication:} 
We perform a Multiplication operation while computing the new spring vector, as shown by Eqn. \ref{eqQ}. Here, the \textit{Diff} tensor and spring force vector are multiplied . This operation is divided into two kernels: Multiplication and Reduction Kernel.
To show the functioning of Multiplication kernel, the first row of \textit{Diff} tensor is multiplied with the spring force vector and the elements are denoted with an alphanumeric sequence as follows. Note that these are elements of the first row and should not be confused as three rows.

\begin{equation}
{Mult}=\begin{bmatrix}
{a1}=({{D}}_{{1000}}-{{D}}_{{0000}}){{F}}_{{1}} & {a2}=({{D}}_{{1001}}-{{D}}_{{0001}}){{F}}_{{2}} & \\ {a3}=({{D}}_{{1002}}-{{D}}_{{0002}}){{F}}_{{3}} &  {a4}=-({{D}}_{{1101}}-{{D}}_{{0100}}){{F}}_{{1}} & \\  {a5}=-({{D}}_{{1101}}-{{D}}_{{0101}}){{F}}_{{2}} & {a6}=-({{D}}_{{1102}}-{{D}}_{{0102}}){{F}}_{{3}} 
\end{bmatrix}
\end{equation}

Thus, upon completion of multiplication, we get the final matrix as shown in Eqn. \ref{EqComMult}: 

\begin{equation}
\textit{Mult}=\begin{bmatrix}
{a1} & {a2} & {a3} &  {a4} & {a5} & {a6} \\
{a7} & {a8} & {a9} &  {a10} & {a11} & {a12} \\
{a13} & {a14} & {a15} &  {a16} & {a17} & {a18} 
\end{bmatrix}
\label{EqComMult}
\end{equation}

This matrix is generated by the Multiplication kernel. 

\textit{Reduction Kernel}:

In this kernel, the rows of the $\textit{Mult}$ matrix are reduced to a single value as shown in Fig. \ref{fig:PictorialRepresentation} and Fig. \ref{fig:algo-reduction}.
\\
\begin{figure}[hbt!]
    \centering
\begin{algorithm}[H]
\tikzstyle{arrow} = [thick,->,>=stealth]
\begin{tikzpicture}[
roundnode/.style={circle, draw=green!60, fill=green!5, very thick, minimum size=7mm},
squarednode/.style={rectangle, draw=red!60, fill=red!5, very thick, minimum size=5mm},
]
\node[roundnode] at (0,1) (pro1) {\color{black}a1};
\node[roundnode] at(1,1) (pro2){\color{black}a2};
\node[roundnode] at (2,1) (pro3){\color{black}a3};
\node[roundnode] at (3,1) (pro4){\color{black}a4};
\node[roundnode] at (4,1)(pro5){\color{black}a5};
\node[roundnode] at (5,1)(pro6){\color{black}a6};
\node[roundnode] at (6,1) (pro13){\color{black}a7};
\node[roundnode] at (7,1) (pro14){\color{black}a8};
\node[roundnode] at (8,1) (pro15){\color{black}a9};
\node[roundnode] at (9,1)(pro16){\color{black}a10};
\node[roundnode] at (10,1)(pro17){\color{black}a11};
\node[roundnode] at (11,1) (pro18){\color{black}a12};
\node[squarednode] at (0.5,-0.5)(pro7){\color{black}b1};
\node[squarednode] at (2.5,-0.5)(pro8){\color{black}b2};
\node[squarednode] at (4.5,-0.5)(pro9){\color{black}b3};
\node[squarednode] at (6.5,-0.5)(pro19){\color{black}b4};
\node[squarednode] at (8.5,-0.5)(pro20){\color{black}b5};
\node[squarednode] at (10.5,-0.5)(pro21){\color{black}b6};
\node[squarednode] at (1.5,-2.5)(pro10){\color{black}b1+\color{black}b2};
\node[squarednode] at (4.5,-2.5)(pro11){\color{black}b3};
\node[squarednode] at (7.5,-2.5)(pro22){\color{black}b4+\color{black}b5};
\node[squarednode] at (10.5,-2.5)(pro23){\color{black}b6};
\node[squarednode] at (3,-4.5)(pro12){\color{black}b1+\color{black}b2+\color{black}b3};
\node[squarednode] at (9,-4.5)(pro24){\color{black}b4+\color{black}b5+\color{black}b6};

\node at (12,-2.5)(ex){\textbf{.....}};
\draw[blue, very thick](-0.5,1.5)--(5.5,1.5);
\draw[blue, very thick] (-0.5,1.5)--(-0.5,-5);
\draw[blue, very thick] (-0.5,-5)--(5.5,-5);
\draw[blue, very thick] (5.5,-5)--(5.5,1.5);
\draw[blue, very thick](5.5,1.5)--(11.5,1.5);
\draw[blue, very thick](5.5,-5)--(11.5,-5);
\draw[blue, very thick](11.5,-5)--(11.5,1.5);
\node at (3,-5.5)(cb){\color{black}\textbf{CUDA-Block-0}};
\node at (9,-5.5)(cb){\color{black}\textbf{CUDA-Block-1}};
\draw[arrow] (pro1)--(pro7);
\draw[arrow] (pro2)--(pro7);
\draw[arrow] (pro3)--(pro8);
\draw[arrow] (pro4)--(pro8);
\draw[arrow] (pro5)--(pro9);
\draw[arrow] (pro6)--(pro9);
\draw[arrow] (pro7)--(pro10);
\draw[arrow] (pro8)--(pro10);
\draw[arrow] (pro9)--(pro11);
\draw[arrow] (pro10)--(pro12);
\draw[arrow] (pro11)--(pro12);
\draw[arrow] (pro13)--(pro19);
\draw[arrow] (pro14)--(pro19);
\draw[arrow] (pro15)--(pro20);
\draw[arrow] (pro16)--(pro20);
\draw[arrow] (pro17)--(pro21);
\draw[arrow] (pro18)--(pro21);
\draw[arrow] (pro19)--(pro22);
\draw[arrow] (pro20)--(pro22);
\draw[arrow] (pro21)--(pro23);
\draw[arrow] (pro22)--(pro24);
\draw[arrow] (pro23)--(pro24);
\end{tikzpicture}
\caption{Reduction Algorithm}
\end{algorithm}
 \caption{Pictorial representation of the Reduction Algorithm.}
    \label{fig:PictorialRepresentation}
\end{figure}
\begin{figure}[hbt!]
    \centering
\begin{algorithm}[H]
    \DontPrintSemicolon
 \vspace{1 mm}
    \SetAlgoLined
    \vspace{1mm}
    \SetKwFunction{FMain}{Reduction}
    \SetKwProg{Fn}{subroutine}{}{}
        \Fn{\FMain}
     {
         declare an extern shared array sdata[]\\
         declare tid = threadIdx.x\\
         declare idx = blockDim.x*blockIdx.x+threadIdx.x
         \\
         sdata[tid]=Mult[idx]\\
          $\_\_$syncthreads()
          \\
           \For{$s \leftarrow 1$ \KwTo $s < Columns$} 
            { 
               declare index=2*s*tid
               \\
               \vspace{1mm}
                \If{(($index < Columns$) $\&\&$ ($(index+s)<Columns$))}
                 {
                   sdata[index]=sdata[index]+sdata[index+s]
                 }
          $\_\_$syncthreads()
          \\
          $s=s*2$
        }
                      
    }
        
    \vspace{1mm}
    \textbf{end subroutine}
    \vspace{1mm}
    \caption{Reduction Algorithm}
\end{algorithm}
\caption{Parallel pseudocode for the Reduction Algorithm.}
    \label{fig:algo-reduction}
\end{figure}
Thus, in the Multiplication kernel, the \textit{Diff} tensors made out of the diffusion and the sigma tensors are multiplied with the spring force and random vectors, respectively, and the results are stored in arrays. Then, in the Reduction kernel, the rows of these arrays are reduced by using the reduction algorithm.

\textbf{Generation of Random Numbers:}
The random numbers are generated for computing the Brownian forces and generating different initial configurations of the chain. For this, we have used the function \textit{curandGenerateUniformDouble} from the GPU library \textit{cuRAND}.
\section{Results and Discussion}
\begin{figure}[hbt!]
\centering
\includegraphics[scale=0.5]{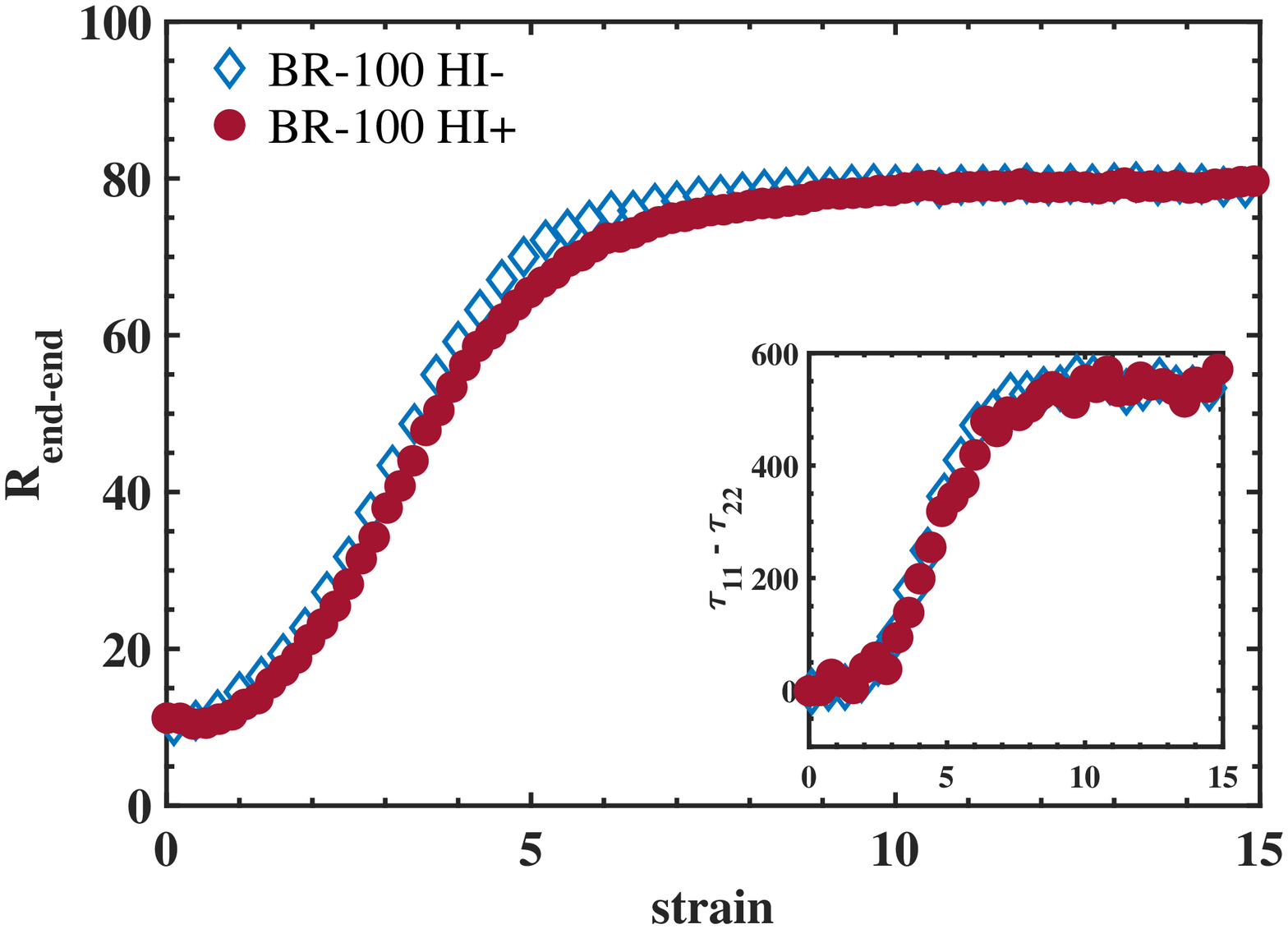}
\caption{The temporal variation of the average end-to-end distance ($R_{end-end}$) of a polymer chain with strain for an uniaxial extensional flow of \textit{Wi}= 3. BR-100 denotes a bead-rod chain with 100 Kuhn steps. HI+ implies that HI is present and HI- implies that HI is absent in the simulations. For the simulations with HI, the value of $h^{*}$=0.00001. In the inset, the first normal stress difference $N_{1}$ for the bead-rod chain, with and without HI, is shown. This is a further check on the GPU implementation used here, particularly in flow fields. When h* is nonzero, but negligibly low, the results are expected to match those without HI. Our results clearly show this for a chain of 100 Kuhn steps, for the extensional flow considered here.}
 \label{fig:Complowhstar}
\end{figure}

First, we test the accuracy of the GPU implementation of BD simulations of a bead-rod chain, written using CUDA C, and check its performance on multiple GPU architectures like Volta and Ampere. For this, we perform simulations in the absence  of a flow field, with HI being active. The results are benchmarked against known equilibrium scaling laws. For further validation, we performed one BD simulation of a chain of 100 Kuhn steps with an extremely low value of h* in an uniaxial extension at flow (shown in Fig. \ref{fig:Complowhstar}). For such negligibly low h*, the results are expected to agree with the same simulations without HI. This is evident from the results presented in Fig. \ref{fig:Complowhstar}, thereby confirming the accuracy of our GPU implementation. Next, when we deal with flow simulations, the bead-rod model predictions are compared with the bead-spring model, with and without HI. Here, we define the chain properties that will be calculated.

\textit{End-to-end distance}: This is the magnitude of the vector joining the terminal beads of the polymer chain. This is calculated as the square root of the ensemble average, as defined elsewhere \cite{larson2005rheology}. This is given as:
\begin{equation}
\mathit{{R}_{end-end}}=\sqrt{\langle(\vec{{r}}_{N}^{\ast}-\vec{{r}}_{0}^{\ast})^2\rangle}
\end{equation}
where "$\langle...\rangle$" denotes an ensemble average.

The component of the end-to-end distance along the y-direction is given by:
\begin{equation}
\mathit{{R}_{end-end,y}}=\sqrt{\langle(\vec{{y}}_{N}^{\ast}-\vec{{y}}_{0}^{\ast})^2\rangle}
\end{equation}
The same in the other directions are defined analogously.

\textit{Radius of gyration}:
This is another measure of the chain size, used in previous studies \cite{larson2005rheology}. This provides an idea of the distribution of beads around the center of mass and is defined as:
\begin{equation}
\mathit{{R}_{g}}=\sqrt{\frac{\sum_{i=0}^N\langle(\vec{{r}}_{i}^{\ast}-\mathit{\vec{{r}}_{CM}^{\ast}})^2\rangle}{{N+1}}}
\end{equation}
where $\vec{{r}}_{CM}^{\ast}$ denotes the center of mass of the chain.

\textit{Relaxation time} ($\lambda$): This is the time scale of the decay of the autocorrelation function of the end-to-end vector of the chain, as defined in an earlier study \cite{dalal2012multiple}. This is obtained by fitting the last 70\% of the auto-correlation function to an exponential decay.

\textit{Weissenberg number}:
It is a dimensionless number \cite{textbook}, defined as the product of the shear/extensional rate and the relaxation time of the polymer chain.

\textit{strain}:
It is the dimensionless product of the simulation time and the shear/extensional rate.

\textit{stress}:
The dimensionless polymeric contribution to the stress  $\overleftrightarrow{\tau}$ (normalized by $n_{p}k_{B}T$, where $n_{p}$  is the number density of polymer chains), is calculated using the Kramer’s expression \cite{osti_6164599}, which is given by:
\begin{equation}
\mathit{\overleftrightarrow{\tau}}=N\overleftrightarrow{\delta}+{\sum_{i=0}^N\langle\vec{{F}}_{i}^{S \ast}\mathit{\vec{{Q}}_{i}^{\ast}}\rangle}
\end{equation}
where $N\overleftrightarrow{\delta}$ is the isotropic part of polymer contribution to the stress tensor, $\vec{{F}}_{i}^{S \ast}$ is the spring force acting on the $i^{th}$ spring and $\vec{{Q}}_{i}^{\ast}$ is the link vector. The first normal stress difference is given by \cite{osti_6164599}:
\begin{equation}
\mathit{N_{1}}={\overleftrightarrow{\tau}_{11}-\overleftrightarrow{\tau}_{22}}
\end{equation}

\begin{figure}[hbt!]
\centering
\includegraphics[scale=0.5]{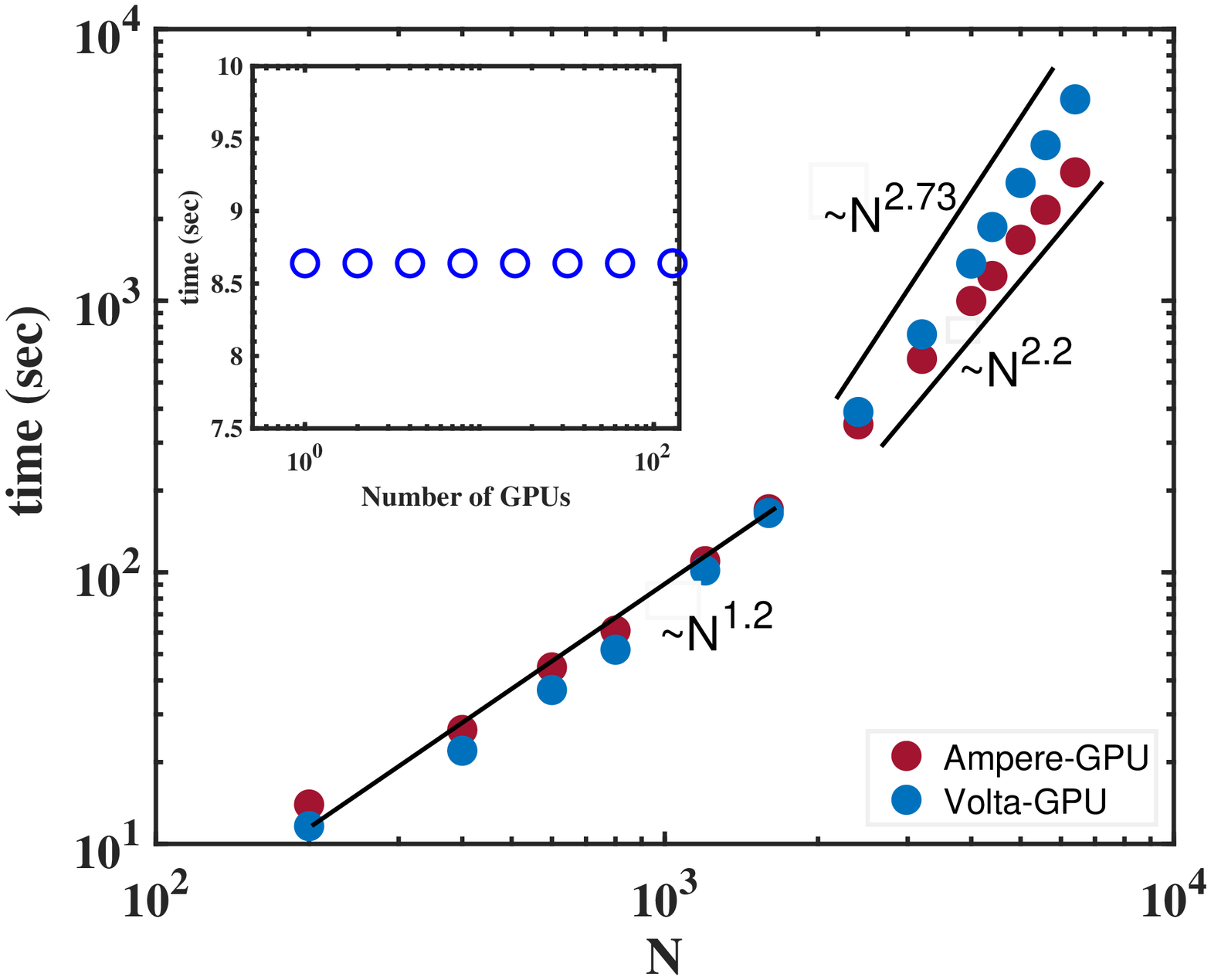}
\caption{Computational time variations with respect to N on different GPU Architectures, for 10,000 time steps of the BD simulation code. In the legend, Volta-GPU represents NVIDIA V100-SXM2-16GB and Ampere-GPU represents NVIDIA A100-80GB. The computational time decreases as we move to new architectures. In the inset, we show the computational time variation with increasing number of GPUs. The times are similar, since there is no intercommunication between different MPI processes. The GPU used is Telsa V100-PCIE-32GB.}
\label{VA100}
\end{figure}

\textbf{Performance on single GPU and scaling across nodes}:
We test the performance of the code on two latest GPU architectures by comparing the time taken to compute 10,000 iterations for different values of $N$, ranging from 199 to 6399, in the absence of flow. 
These computational times are shown in Fig. \ref{VA100}.

From Figure \ref{VA100}, we observe that, as the number of springs of the chain reaches $N$=4399, the computational time on the A100 becomes significantly lower than those on V100.  The computational time is almost half at $N$=6399. This analysis shows the effect of the GPU generation on the computational time, thereby justifying the usage of modern GPUs. 
Accordingly, the scaling also reduces from $N^{2.73}$ on V100 to $N^{2.2}$ on A100. Thus, we obtain a scaling of about $N^{1.2}$ and $N^{2.2}$ for the computational times for smaller and larger chains, respectively. Significantly, these are obtained when
HI is calculated with the conventional algorithm that uses a Cholesky decomposition, which shows a scaling of $N^3$ on a CPU. Interestingly, we are able to obtain a significantly lower exponent without using any of the recently developed methods, just by performing our calculations on a modern GPU. Note that there are other algorithms, developed over the past two decades or so \cite{fixman1986construction,ando2012krylov,geyer2009n} that help reduce the scaling exponent. In this study, we have used the conventional algorithm and yet obtained a similarly large reduction in the scaling exponent solely by processing on a modern GPU.

\textit{Scaling across the nodes:}
 To find the scaling of the hybrid code across multi-nodes and multi-GPUs, we have performed a flow simulation of a 100 bead-rod chain for 20,000 iterations at $\textit{Wi}$ = 3 on 1,2,4,8,..128 GPUs. The results shown in the inset of Fig. \ref{VA100} shows a strong scaling behaviour across the GPUs as expected, since there is no inter-communication between the different MPI processes.

\begin{figure}[hbt!]
\centering
\includegraphics[scale=0.5]{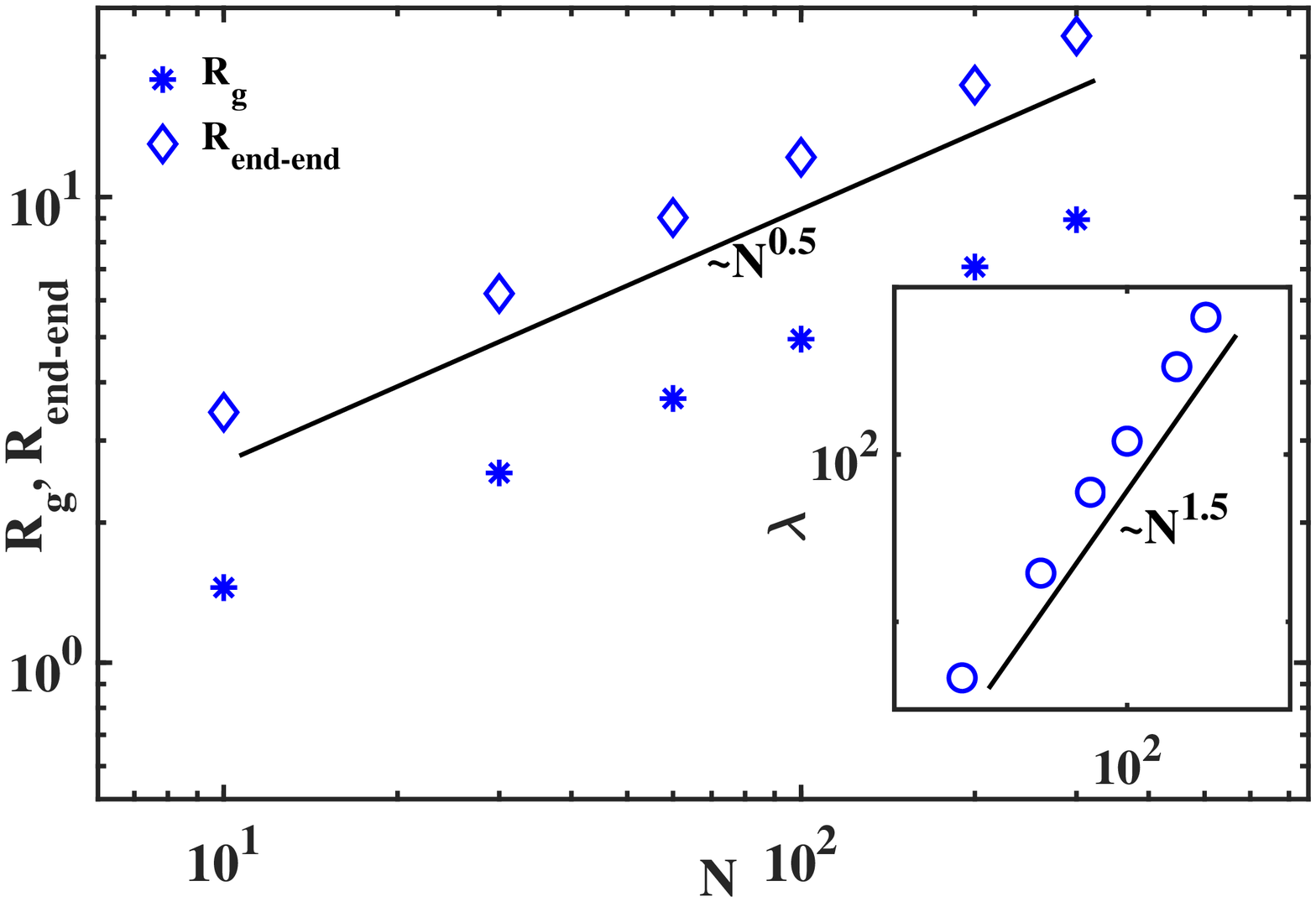}
    \caption{The variation of $R_{end-end}$, $R_{g}$ and $\lambda$ with $N$ (number of beads), at equilibrium. HI is active for these BD simulations. The GPU used is NVIDIA V100 -SXM2-16GB. The scaling exponents obtained are in perfect agreement with theoretical expectations.}
    \label{fig:Eqbrendrogtau}
\end{figure}

\textbf{BD simulations at equilibrium}: These simulations are performed on a single GPU with a time step size of 0.001, stiffness constant of 1000 for the Fraenkel spring and a total run time of 100 relaxation times, for any given chain size. The scaling (with respect to $N$) of the end-to-end distance, radius of gyration and the relaxation time are shown in Fig. \ref{fig:Eqbrendrogtau}. The scaling exponent is observed to be 0.5 for the chain size measures and 1.5 for the relaxation time, in accordance with theoretical expectations \cite{textbook}, which further validates the code. 

\textbf{Petascale BD simulations with flow}:
In the flow simulations, all properties are  averaged over multiple simulations. Unless otherwise mentioned, we average any property over 100 and 1000 independent BD simulations (with different initial configurations and
Brownian noise history), for extensional and shear flows, respectively. Thus, overall, all our results come from Petascale BD simulations on the GPU. The flow simulations using the bead-rod model without HI and bead-spring simulations with and without HI are run on a CPU. When HI is active for the bead-rod simulations, some simulations are performed with two different time step sizes $\Delta t^{*}$=0.001, 0.0001 and spring constants: $K^{*}$=1000, 10000, respectively, to check the consistency of the results. The value of parameters $h^{*}$ and $a^{*}$ in the Eqns \ref{Diffusioneq1}-\ref{Diffusioneq3} are defined for the case of  bead-rod simulations as follows: $h^{*}$=0.5 and $a^{*}$=0.5. This implies that the hydrodynamic bead diameter is equal to one Kuhn step (i.e. one Fraenkel spring in the chain). This further implies that the fully stretched chain would hydrodynamically behave similar to a cylinder with a diameter equal to the bead
diameter. This makes it easier to implement the “matching condition” in the procedure discussed by Hsieh et al. \cite{hsieh2003modeling} to obtain equivalent
parameters for the corresponding bead-spring simulations. Following the calculations, the following values are obtained: h* = 0.105 and a* = 1.539, for a bead-spring chain of 20 springs.

\textit{Extension flow simulation}: 
 For startup extensional flows, we consider a flow rate of \textit{Wi}= 2.5 and 3. For each case, we have averaged $R_{end-end}$ over 128 cases, resulting in a Petascale simulation. These results of averaged $R_{end-end}$ are compared with a bead-rod chain without HI and the equivalent bead-spring chain with and without HI. 
 
\begin{figure}[hbt!]
\centering
{\includegraphics[scale=0.5]{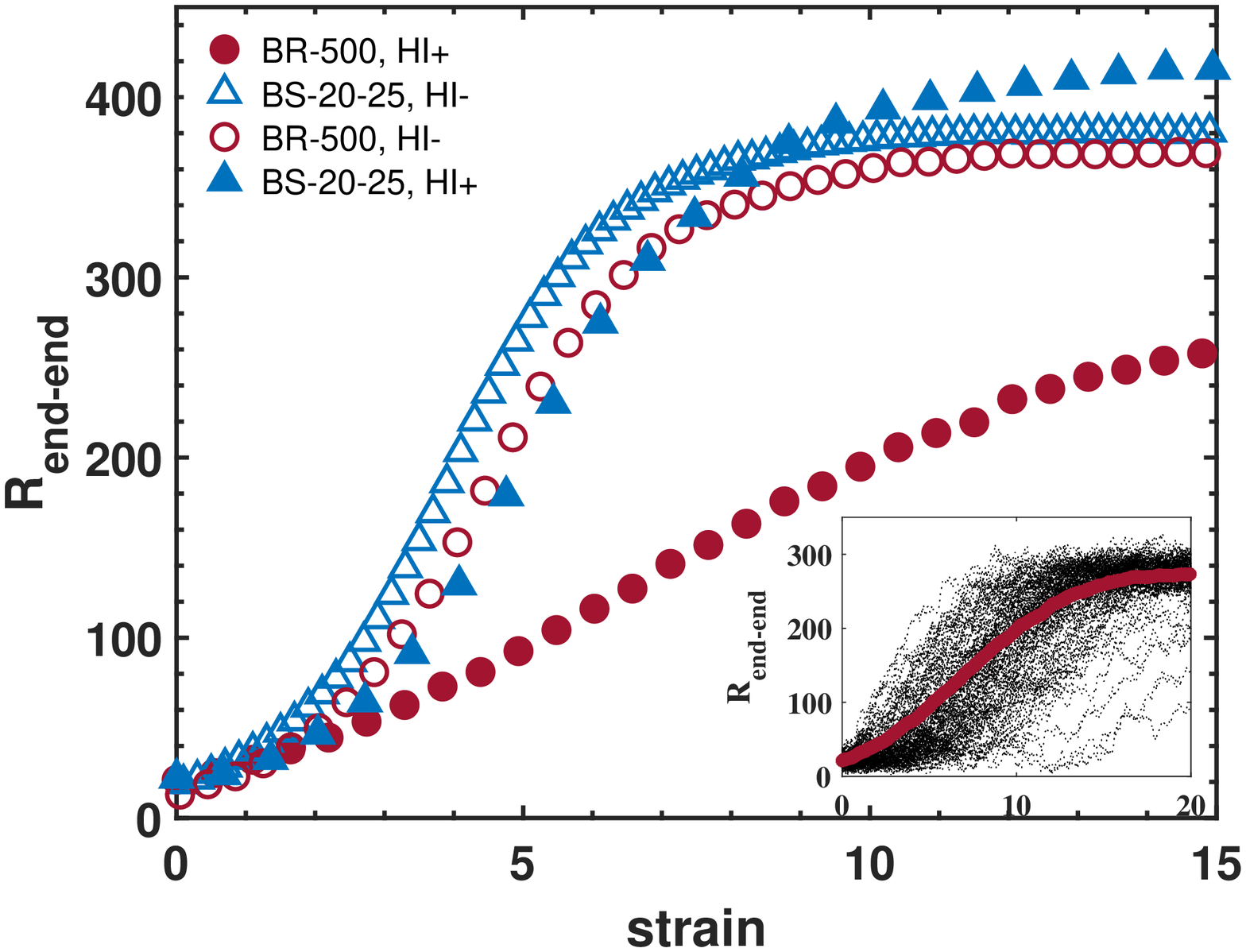} }
\caption{The average variation of the end-to-end distance ($R_{end-end}$) of polymer chain with strain in an uniaxial extensional flow of \textit{Wi}= 2.5. BR-500 denotes a bead-rod chain with 500 Kuhn steps and BS-20-25 denotes a bead-spring chain having 20 springs, with 25 Kuhn steps mimicked by each spring. HI+ implies that HI is present and HI- implies that HI is absent in the simulations. The Cohen-Padé approximation is used as the spring law for the bead-spring models. The results shown here are averaged over 128 cases. We have used 16 DGX's, each having 8 GPUs of Tesla V100-PCIE-32GB. The CPU used is Intel Core-i7@3.4 GHz and 32 GB RAM. The inset shows the behavior of the entire ensemble and the average for the bead-rod chain of 500 Kuhn steps, with HI being active.} 
\label{fig:HIWHIWi2.5}
\end{figure}

\begin{figure}[hbt!]
\centering
\includegraphics[scale = .5]{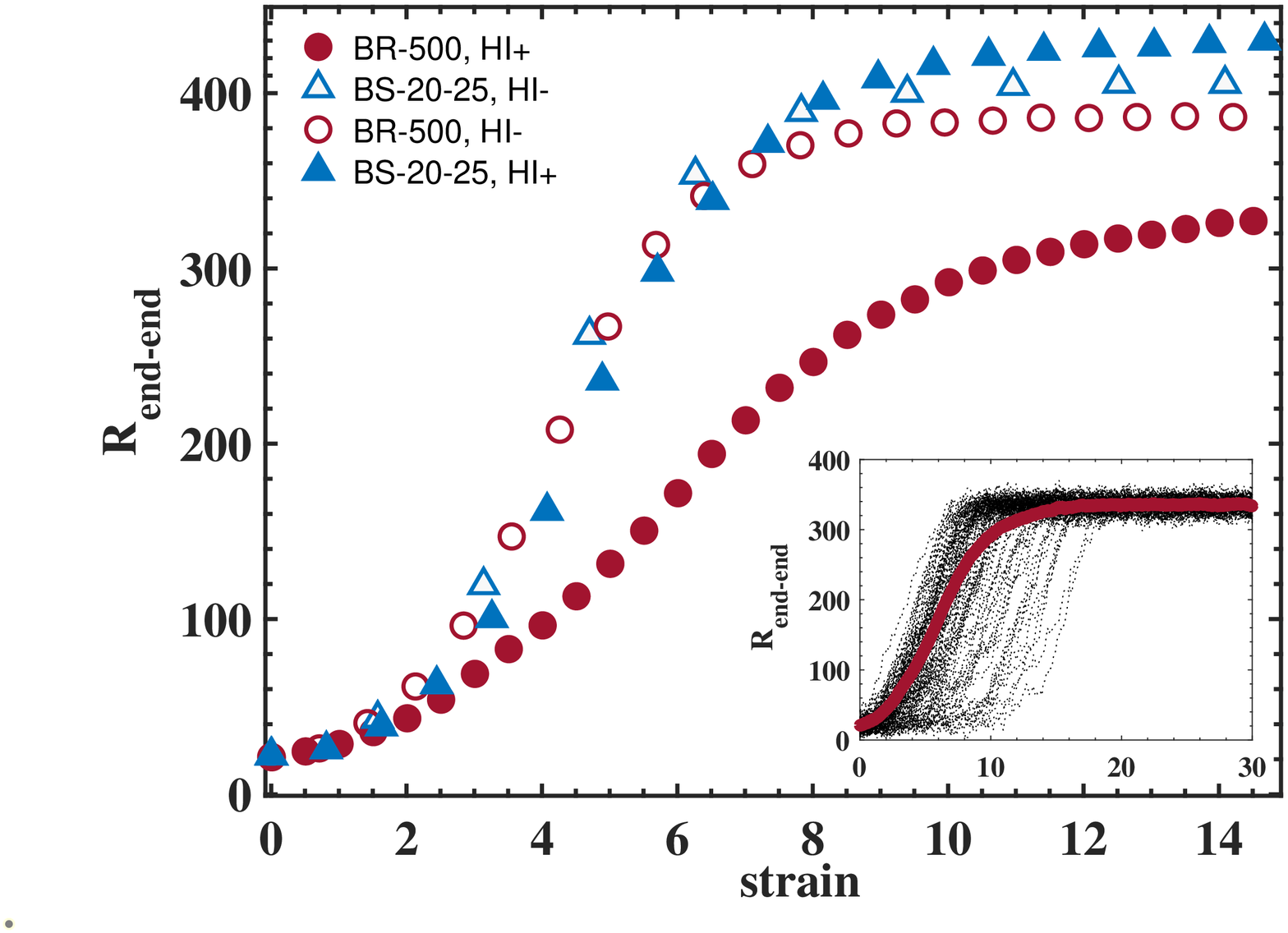}
 \caption{Same as Fig. \ref{fig:HIWHIWi2.5}, except for \textit{Wi}= 3.}
\label{fig:HIWHIWi3}
\end{figure}

\begin{figure}[hbt!]
\centering
\includegraphics[scale=0.95]{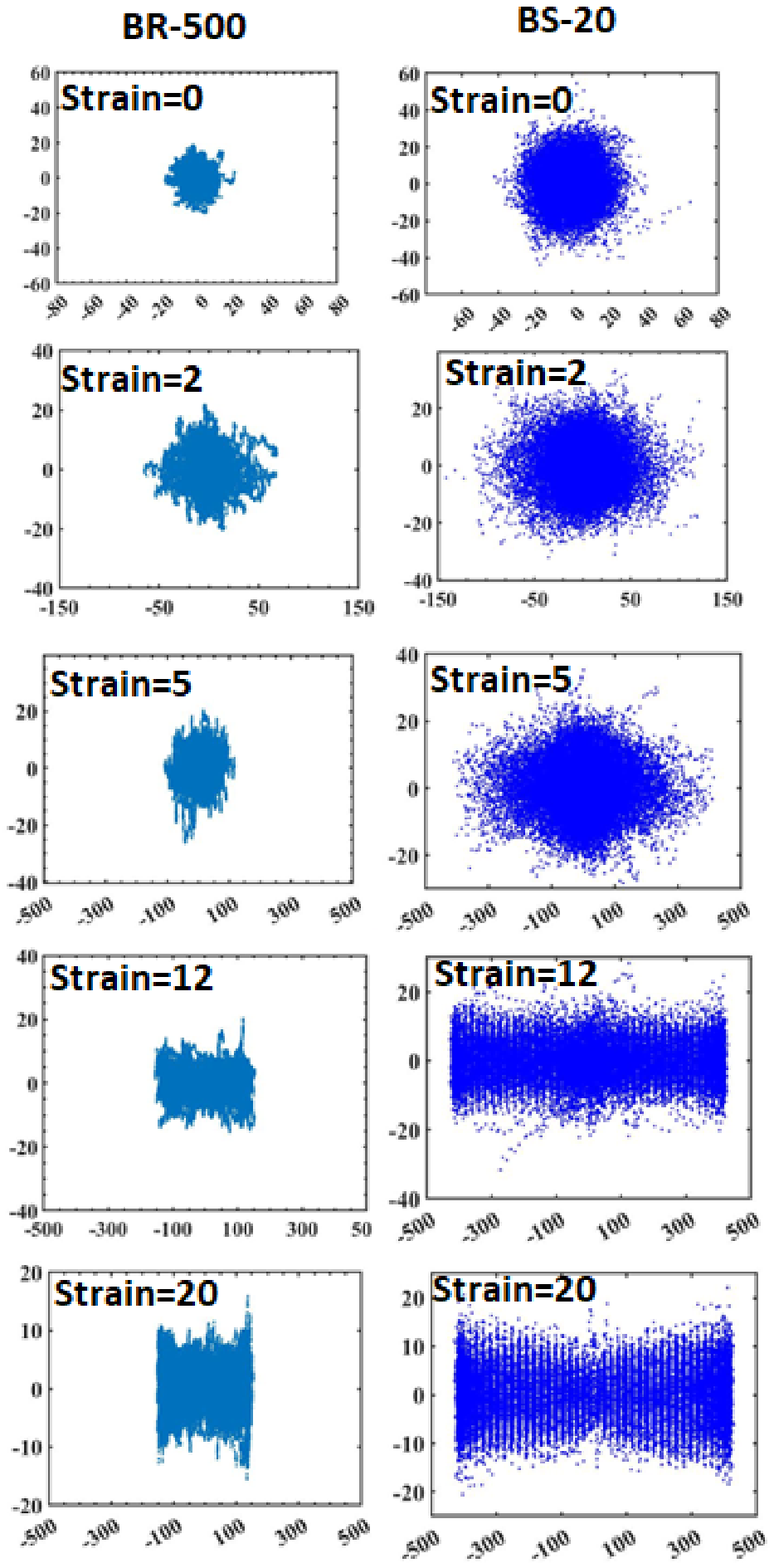}
 \caption{The temporal evolution of the entire ensemble for \textit{Wi}= 2.5 for the bead-rod (left) and bead-spring (right) models. The strain (dimensionless simulation time) increases from top to bottom in both sets (left and right). Note how the ensemble evolves from a spherical coil (equilibrium) to a roughly cylindrical shape, which is much longer for the bead-spring model.}
\label{fig:TempEvol2.5}
\end{figure}

Figs. \ref{fig:HIWHIWi2.5} and \ref{fig:HIWHIWi3}  show the variation of average $R_{end-end}$ with strain for extensional flows of \textit{Wi}= 2.5 and 3, respectively, for both bead-rod and bead-spring models in the absence and presence of HI. Note, all chain models considered have the same number of Kuhn steps and are exposed to a flow of similar strength (with respect to \textit{Wi}). Without HI, the bead-spring model is constructed such that the total number of Kuhn steps are same, with a reasonably large number of springs. With HI, the parameters were calculated by the procedure mentioned in the study by Hsieh et al. \cite{hsieh2003modeling}, as mentioned earlier. The inset to both Figs. \ref{fig:HIWHIWi2.5} and \ref{fig:HIWHIWi3} show the stretch of all individual cases (dotted black lines) and the ensemble average (which is shown in the main Figure). These highlight the variability in the temporal behavior within the ensemble. Further, to contrast the chain stretch evolution from the models in the presence of HI, the temporal evolution of the entire ensemble is shown in Fig. \ref{fig:TempEvol2.5}. For this, at any strain, all beads on all chains
in the ensemble are shown as dots. When observed together, they indicate the average stretch of the ensemble at that strain. The results presented in Figs. \ref{fig:HIWHIWi2.5}-\ref{fig:TempEvol2.5} clearly highlight the impact of using highly resolved chains, especially when HI is active. From an earlier study \cite{https://doi.org/10.48550/arxiv.2208.04457}, it is known that the bead-spring model predictions for stretch are largely consistent with those of the bead-rod model, in the absence of HI. However, a significant inconsistency was observed in the stress predictions. Accordingly, results shown here also agree well across bead-spring and bead-rod models, when HI is not present. However, there is a very significant offset, even in the stretch predictions, when HI is active. This clearly shows the importance of using highly resolved models, especially for long chains in the presence of HI. Such large differences in stretch predictions is bound to produce significant inconsistencies in stress predictions, which is further expected to affect any predictions of the resulting flow field for any practical applications involving dilute polymer solutions.


\begin{figure}[hbt!]
\centering
\includegraphics[scale=0.5]{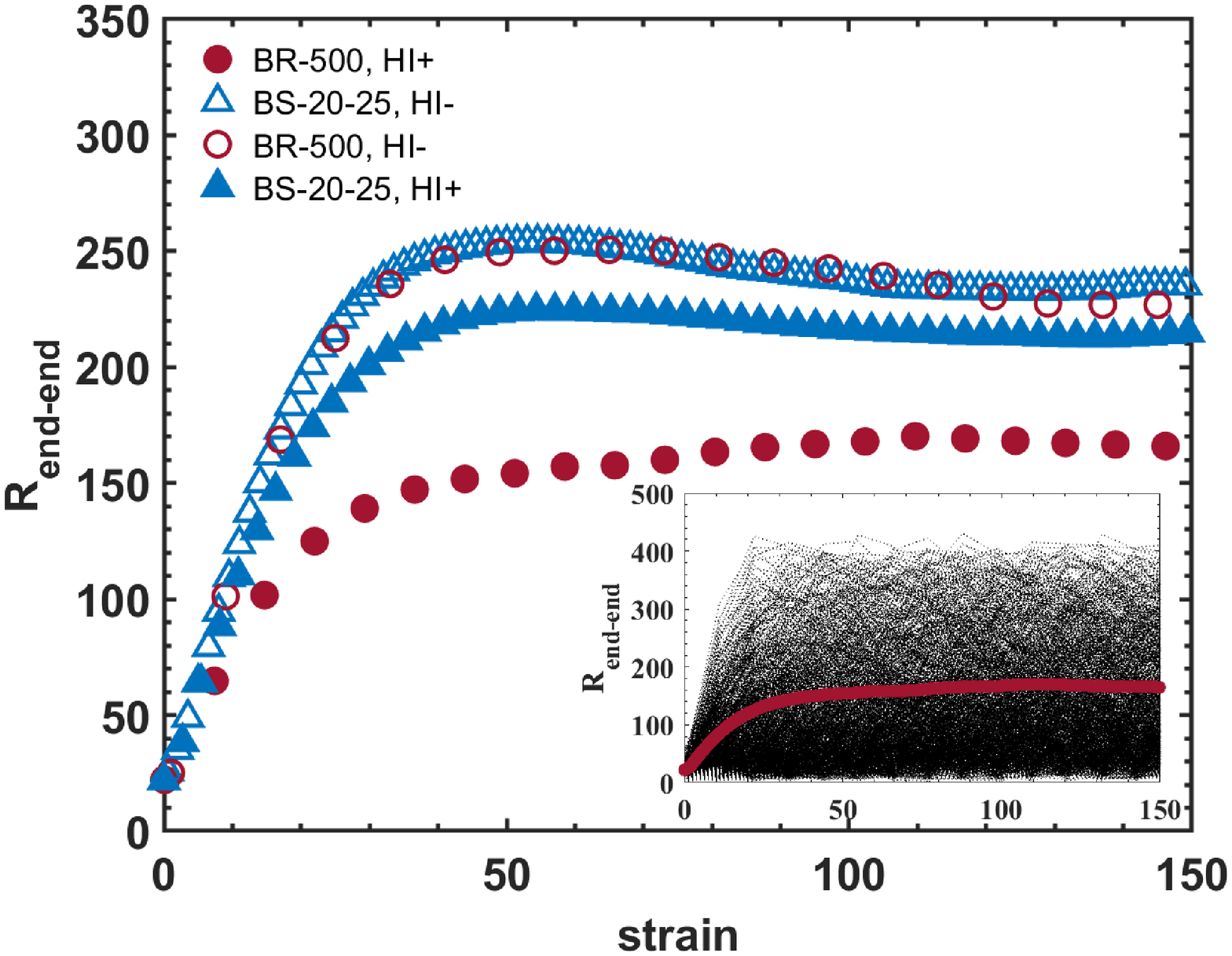}
\caption{The average variation of the end-to-end distance ($R_{end-end}$) of polymer chain with strain in a steady shear flow of \textit{Wi}=100. BR-500 denotes a bead-rod chain with 500 Kuhn steps and BS-20-25 denotes a bead-spring chain having 20 springs, with 25 Kuhn steps mimicked by each each spring. HI+ implies that HI is present and HI- implies that HI is absent in the simulations. The Cohen-Padé approximation is used as the spring law for the bead-spring models. The results shown here are averaged over 1000 cases.We have used 16 DGX's, each having 8 GPUs of Tesla V100-PCIE-32GB.The CPU used is Intel Core-i7@3.4 GHz and 32 GB RAM. The inset shows the behavior of the entire ensemble and the average for the bead-rod chain of 500 Kuhn steps, with HI being active.}
 \label{fig:BRBSWi100HI}
\end{figure}

\begin{figure}[hbt!]
\centering
\includegraphics[scale=0.5]{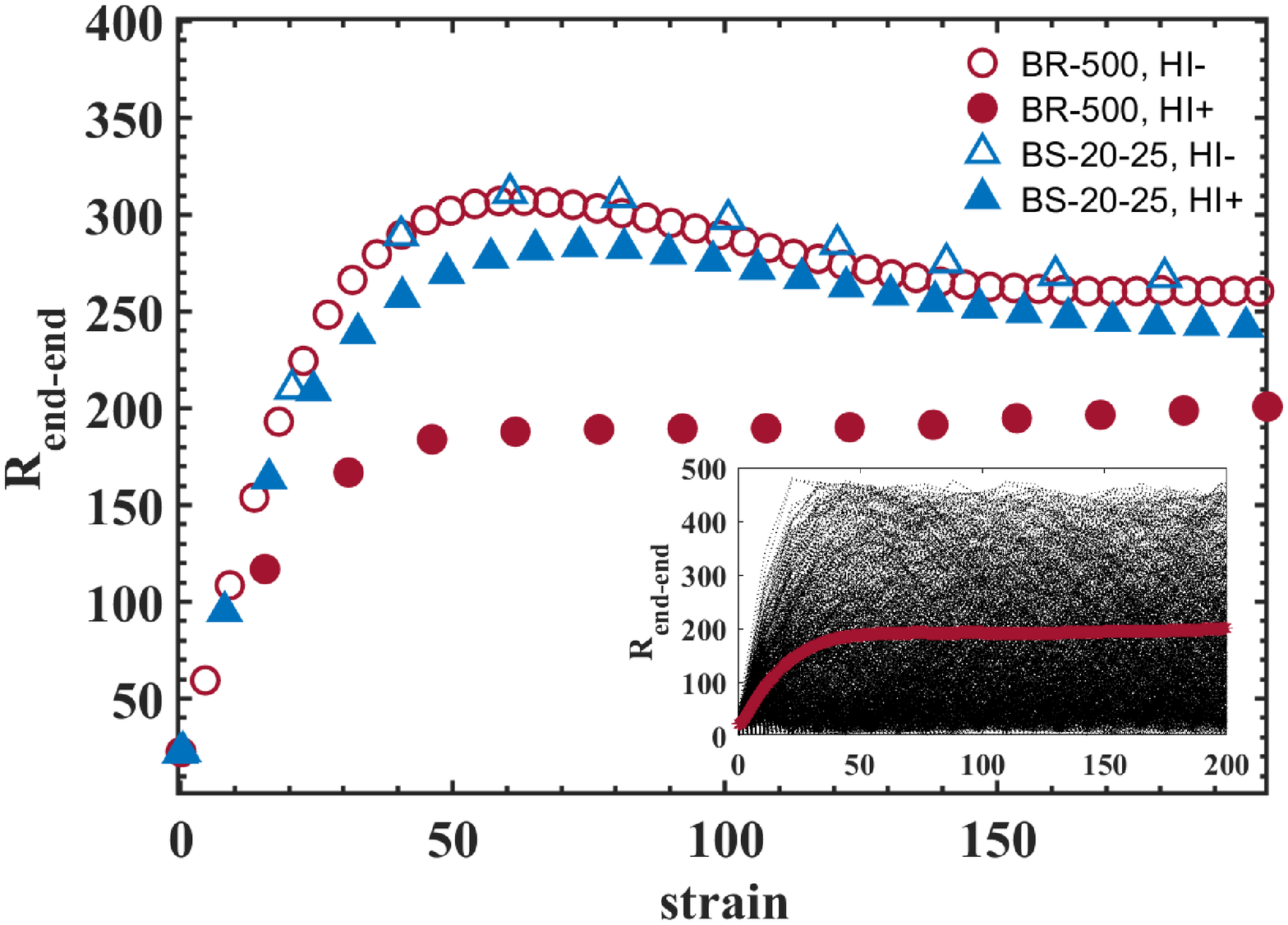}
\caption{Same as Fig. \ref{fig:BRBSWi100HI}, except for \textit{Wi}=300.}
 \label{fig:Wi300HI}
\end{figure}

\begin{figure}[hbt!]
\centering
\includegraphics[scale=1]{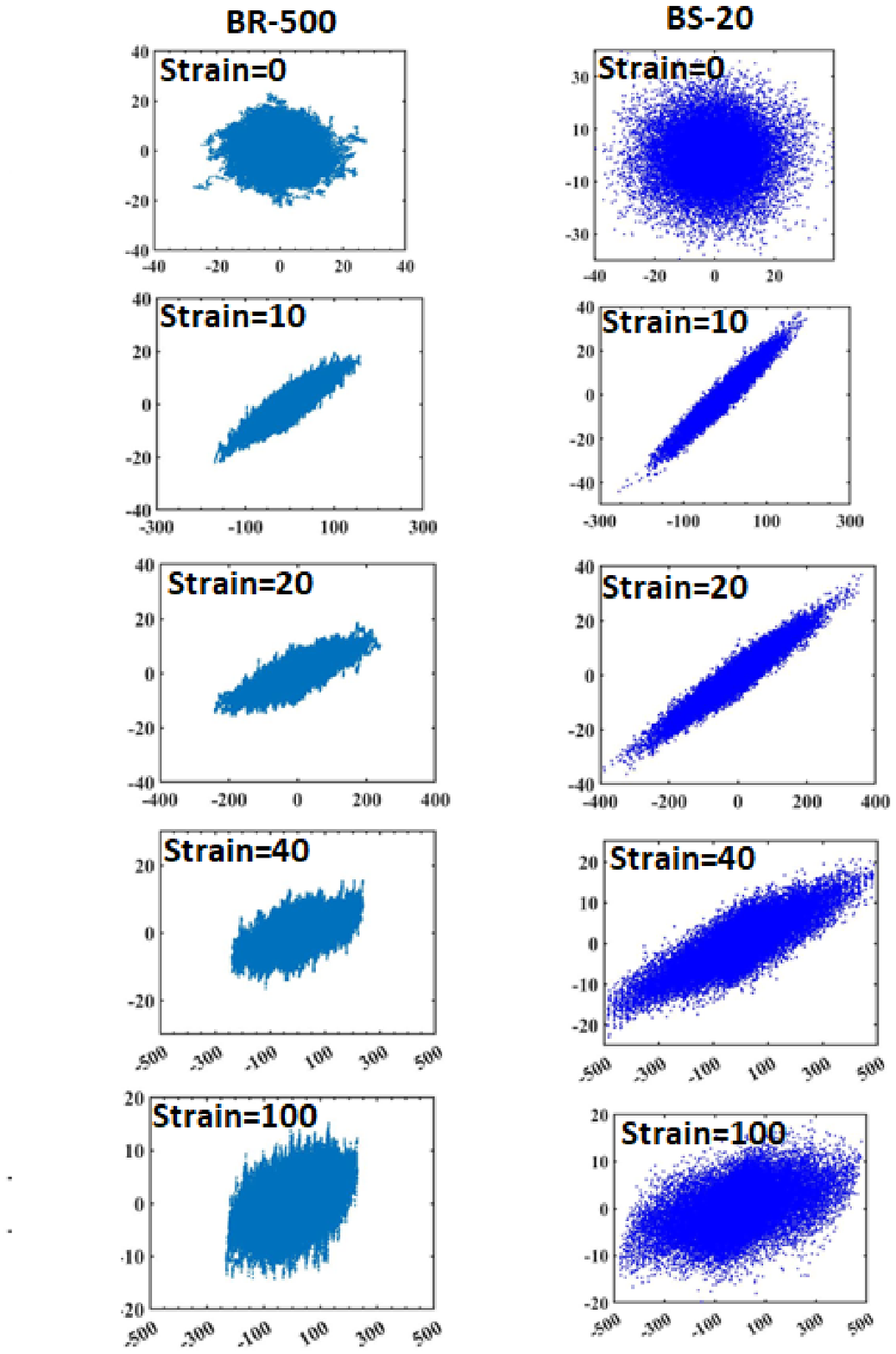}
\caption{Same as  Fig. \ref{fig:TempEvol2.5}, except for steady shear flow of \textit{Wi} = 300. Note how the ensemble evolves from a spherical coil (equilibrium) to a roughly inclined cigar shape, which is much longer for the bead-spring model.}
 \label{fig:TempEvol300}
\end{figure}

\textit{Shear flow simulation}: For startup shear flows, we consider flow rates of \textit{Wi}= 100 and 300. The ensemble consisted of 1000 cases, over which all results are averaged. Thus, to completely analyze at any shear rate, we have used multiple Petascale BD simulations. Figs. \ref{fig:BRBSWi100HI} and \ref{fig:Wi300HI} show the variation of average $R_{end-end}$ with strain for steady shear flows of \textit{Wi} = 100 and 300, respectively, for both bead-rod and bead spring models in the absence and presence of HI. The parameters used here are similar to those for the extensional flow simulations. The insets to both Figs. \ref{fig:BRBSWi100HI}  and \ref{fig:Wi300HI}  show the stretch of all individual cases (dotted black lines) and the ensemble average (which is shown in the main Figure). Further, to contrast the chain stretch evolution from the models in the presence of HI, the temporal evolution of the entire ensemble is shown in Fig. \ref{fig:TempEvol300} for \textit{Wi} = 300. The overall trends obtained are similar to those of the extensional flow simulations. The results are consistent across bead-spring and bead-rod models, when HI is not present. The bead-spring results are only slightly affected when HI is active. However, there is a very significant offset in the predictions of the final steady state stretch, when HI is active. Even more significant is the absence of the overshoot in the bead-rod stretch for the flow rates considered, when HI is active, which is in sharp contrast to the behavior obtained without HI as well as with the bead-spring model, with HI. The shear flow results further highlight the impact of using such highly resolved models, especially for long chains in the presence of HI, even though the computational requirements are much higher.

\begin{figure}[hbt!]
\centering
\includegraphics[scale=0.5]{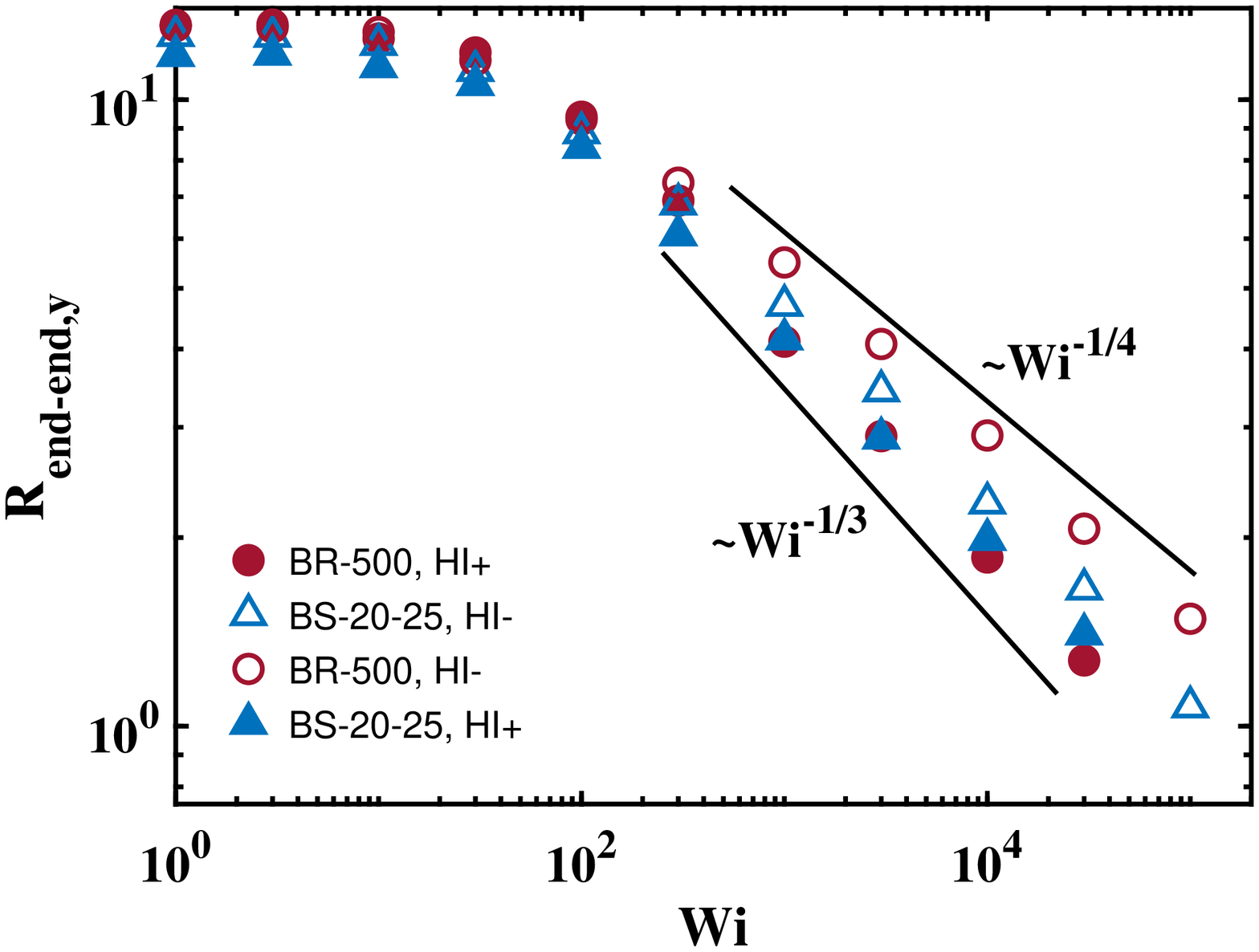}
\caption{Variation of the steady state chain thickness ($R_{end-end,y}$) of a polymer chain with \textit{Wi} in shear flow. All legends have similar meaning as Fig. \ref{fig:BRBSWi100HI}. Note that the bead-rod model follows $\sim \textit{Wi}^{-1/4}$ and $\sim \textit{Wi}^{-1/3}$ in the absence and presence of HI, respectively, confirming the theoretical analysis presented in an earlier study \cite{doi:10.1021/ma3014349}. In contrast, the bead-spring models always show a scaling of $\sim \textit{Wi}^{-1/3}$, irrespective of whether HI is active or absent.}
 \label{fig:RendyVsWi}
\end{figure}

With respect to shear flows, another interesting quantity is the steady state chain thickness, and the scaling law of the same, obtained from different models. Earlier studies \cite{doi:10.1021/ma3014349} indicate a scaling law of -1/4 and -1/3 for bead-rod and bead-spring models, respectively, in the absence of HI. The analysis presented in the study by Saha Dalal et al. \cite{doi:10.1021/ma3014349} hints at a -1/3 scaling law for longer chains using the bead-rod model, when HI is dominant. However, till date, this possibility is not confirmed from BD simulations. This is shown for the first time in this study, since we are able to use long, highly resolved chains, with HI. The relevant results are presented in Fig. \ref{fig:RendyVsWi}. Clearly, the chain thickness obtained from the bead-rod model $\sim\textit{Wi}^{-1/3}$ when HI is active and $\sim\textit{Wi}^{-1/4}$ when HI is absent, whereas the bead-spring model always shows an exponent of -1/3. This further highlights the inadequacy of such bead-spring formulations as well as confirms the analysis presented in the earlier study \cite{doi:10.1021/ma3014349}.

\newpage
\textbf{Conclusions}

To summarize, this manuscript presents the details of a scalable, parallel, GPU implementation of BD simulations, incorporating the Cholesky decomposition technique for the inclusion of HI, using a hybrid code. The application is implemented using CUDA and MPI to leverage multiple levels of parallelism exposed by multi-node, multi-GPU systems. As mentioned earlier, such GPU implementations of BD simulations for polymer dynamics on modern GPUs don’t exist, to the best of our knowledge. The earlier ones have neither been tested exhaustively (unlike our study presented here) by users, nor are they compatible with modern day GPUs. The following are the key conclusions obtained from this study:
\begin{itemize}
 \item Our GPU implementation shows a strong scaling behaviour across nodes. Our bead-rod simulations show two different scalings for the computational times - $N^{1.2}$ for short chains when the GPU is not completely saturated and about $N^{2.2}$-$N^{2.8}$, depending on the type of GPU, for longer chains, as the GPU gets closer to saturation. Note, no previous investigation showed the possibility of such lower exponents as 2.2 using the conventional algorithm with the Cholesky decomposition, by using modern GPUs instead of CPUs.
 \item We performed a thorough test for accuracy. Our GPU implementation is able to reproduce the known theoretical scaling laws at equilibrium for both chain size and relaxation time. Further, for very low $h^{*}$, the GPU implementation yields identical results as simulations without HI in flow, as it might be expected.
 \item Our flow results highlight the necessity for using highly resolved chains, to a single Kuhn step, while performing BD simulations in flow.  The bead-rod chain simulations predict much lower stretch for both extensional and shear flows, when compared with the equivalent bead-spring chain at same flow rates. This proves the inadequacy of such coarse-grained bead-spring models in capturing chain dynamics in flows, especially when HI is active. Due to the massive speedup, GPUs are absolutely essential to make such simulations of long, highly resolved chains, practically feasible. 
\end{itemize}
Hence, to further summarize, we are able to perform Petascale BD simulations of polymer chains in flow fields using modern GPU architectures. Not only are we able to obtain better scalings of computational times by using GPUs (in the presence of HI), but also observe a significantly different physics, with lower stretch predictions than conventional bead-spring models, when imposed to different flow fields. Going ahead, we plan to implement the concept of Unified Memory for the simulation of polymer chains with more than 7000 beads.
\newpage
\section{Acknowledgement}
V.S.K and P.K. would like to acknowledge the GPU Hackathon 2019 event organizers and especially the mentors there (Advait Soman and Mayank Jain), who have shared their ideas that helped in the initial development of the GPU code. We gratefully acknowledge the support of NVIDIA AI Technology Center for technology and access of the Tesla V100 GPU used for flow simulations of bead-rod chain with HI. The support and the resources provided by PARAM Sanganak under National Super computing Mission, Government of India at Indian Institute of Technology, Kanpur, are gratefully acknowledged. I.S.D would like to acknowledge the generous support received from the DST SERB project ECR/2017/000085 and the IIT Kanpur initiation grant.

\bibliographystyle{elsarticle-num}
\bibliography{main}

\vskip3pt

\end{document}